\documentclass[twocolumn]{emulateapj}
\usepackage{bm}
\usepackage{amsmath}
\usepackage{amstext}
\usepackage{epsfig}
\usepackage{graphicx}

\shorttitle{uGMRT observations of the Coma cluster}
\shortauthors{Lal, D.V.}

\begin{document}

\title{Upgraded GMRT observations of the Coma cluster of galaxies: The observations}

\author{Dharam V. Lal}
\affil{National Centre for Radio Astrophysics - Tata Institute of Fundamental Research Post Box 3, Ganeshkhind P.O., Pune 41007, India}
\email{dharam@ncra.tifr.res.in}

\begin{abstract}
We have used the upgraded Giant Metrewave Radio Telescope to map the Coma cluster of galaxies at 250--500 MHz and 1050--1450 MHz bands.  These 6\farcs26 and 2\farcs18 resolutions observations allow detailed radio structures to be determined
of all detected radio sources that show both discrete pointlike and extended morphologies.
We present images of a subset of 32 brightest ($S_\nu \gtrsim$ 30~mJy) and dominant sources, and several sources show discrete pointlike radio morphologies.
We find the steepening of the spectra consistent with synchrotron cooling in the majority of sources and the median for spectral indices is $-$0.78, suggesting that $\sim$59\% sources have steep spectra.
The nature and the statistical properties of the radio sources in the Coma cluster will be discussed in subsequent papers. 
\end{abstract}

\keywords{Active galaxies (17); Radio jets (1347); Radio active galactic nuclei (2134); Coma cluster (270); Radio continuum emission (1340); Galaxy clusters (584)}

\section{Introduction}
\label{intro}

The largest hierarchical units of the distribution of matter in the universe, called clusters are asymmetric structures, located at the 
intersecting nodes of the filamentary structure of the cosmic web.
They are constantly evolving and growing by accreting matter from the surrounding large-scale structures.
Beyond radio sources in galaxy clusters, a fraction of merging clusters host cluster-wide diffuse emission, implying the existence of relativistic electrons and a large-scale magnetic field, called the radio halos.  These continue to be the topic of much current research.
Therefore, probing the gaseous medium in galaxy clusters through the study of radio emission of ``halo sources" reveals important information on physical processes in clusters.  Of particular interest is the origin of both the relativistic particles and the associated magnetic field, which are required to generate the observed diffuse synchrotron emission.

The prototype cluster, the Coma cluster of galaxies \citep[Abell\,1656; $z$ = 0.0235][]{1999ApJS..125...35S} contains more than 30 cluster radio galaxies, of different radio morphologies and of different optical morphological types \citep{1990AJ.....99.1381V,1996MNRAS.282..779W,BrownRudnick2011,MHM2009,2015fers.confE..63B}.
Moreover, the Coma cluster is permeated by giant diffuse radio halo emission \citep{Willson1970}, which is not identified with any individual galaxy and makes the Coma cluster exceptional in the radio domain.
In addition, the two dominant galaxies, NGC\,4874 (the brightest cluster galaxy, BCG) and NGC\,4889 have a velocity difference of $\sim$700~km~s$^{-1}$ between them.  Taken together with the presence of radio halo, this suggests that the Coma cluster has undergone a major merger \citep{1987ApJ...317..653F,2005A&A...443...17A,2010ApJ...713..291O}.

\citet{2001A&A...365L..74N} using \textit{XMM}-\textit{Newton} further suggests that NGC\,4839
group, located towards the south is also merging with the Coma cluster.
This nearby cluster is a strong, X-ray bright and hot ($\sim$8 keV) source
\citep[see, e.g.][and references therein]{2013ApJ...775....4S},
which deviates significantly from spherical symmetry \citep{Eckertetal2012},
and is a dynamically active, non-cool core system
\citep{Edgeetal1990,Vikhlinin1999,2001A&A...365L..74N}.

\begin{table*}
\caption{Log of the observations}
\begin{center}
\begin{tabular}{lccccccc}
\hline
\multicolumn{1}{c}{Obs\_band} &  Obs. Date & Cal. & $\Delta\nu$ & No. of Ch. & t$_{\rm int.}$ & FWHM & \textsc{rms}  \\
   & & &  (MHz) &   & (hour) & ($^{\prime\prime}\times^{\prime\prime}, ^{\circ}$)& ($\mu$Jy~beam$^{-1}$) \\
   \multicolumn{1}{c}{(1)} & (2) & (3) & (4) & (5) & (6) & (7) & (8) \\
\hline\noalign{\smallskip}
\multicolumn{8}{l}{uGMRT array} \\
250--500 MHz & 2017 Apr 28 & 3C\,286 & 200 & 4096 & $\sim$1.8 &6.65$\times$5.90, 76.39 & 21.1 \\
1050--1450 MHz & 2017 Apr 26 & 3C\,286 & 400 & 8192 & $\sim$1.5 &2.43$\times$1.95, 69.51 & 12.7 \\
\hline
\end{tabular}
\end{center}
\label{tab:obs-log}
\tablecomments{Col.~8: \textsc{rms} noise at the half power point (see also Sec~\ref{sec.data-red} for a discussion).}
\end{table*}

The sources at the center of the Coma cluster, both the BCG and NGC\,4874, along with the extended radio halo emission are known collectively as Coma~C \citep{Willson1970}.
Here, in this paper, the first in the series, we present observations from the upgraded Giant Metrewave Radio Telescope (uGMRT) project targeted at the enigmatic Coma cluster.  
We aim to make use of the high sensitivity and high resolution of the uGMRT in order to investigate for the first time the detailed radio structures of radio sources in the Coma cluster.  Our images have the advantage of high positional accuracy and sensitivity at low radio frequencies.

We assume a $\Lambda$CDM cosmology with $\Omega_{\rm m}$ = 0.27, $\Omega_{\Lambda}$ = 0.73, and $H_0$ = 70 km s$^{-1}$ Mpc$^{-1}$.
At the redshift of the cluster, 1$^{\prime}$ corresponds to 28 kpc.  Throughout, positions are given in J2000 coordinates.
We define spectral index, $\alpha$, as $S_\nu$ $\propto$ $\nu^\alpha$; where $S_\nu$ is the flux density at frequency, $\nu$.
We present our observations and data analysis in Sec.~\ref{sec.obser} and Sec.~\ref{sec.data-red}, respectively.
We present our observational results in Sec.~\ref{sec.morph-spec}, and we also discuss radio spectra and luminosities (Sec.~\ref{rad-spec}).  Sec.~\ref{sec.sum-conc} summarizes our conclusions and future directions.

\section{Observations}
\label{sec.obser}

The GMRT \citep{Swarupetal1991} has been recently upgraded with a completely new set of receivers at frequencies $<$~1500 MHz.  This upgrade provides the telescope (nearly) seamless frequency coverage from 50 to 1500 MHz \citep{Guptaetal2017}.

The Coma cluster of galaxies was re-observed (proposal ddtb270) at band-3 (250--500 MHz band) and band-5 (1050--1450 MHz) of the uGMRT because of 
increased bandwidth (and hence enhanced ($u,v$) coverage), higher antenna gain, and a lower system temperature of receivers, which all contribute to much improved sensitivity.  This enables it to image high resolution structures along with diffuse, low surface brightness extended structures with good angular resolution and with good sensitivity.
The time sampling of the data was 2.67~s, which is sufficient to sample the phase fluctuations of the ionosphere and to avoid time smearing for sources at the outer edge of the field of view.
The GMRT Wide-band Backend was used as the correlator, spanning the frequency range of 300--500 MHz and 1050--1450 MHz at the 250--500 MHz band and the 1050--1450 MHz band, respectively in two polarizations, RR and LL.
A summary of the observational setup is detailed in Table~\ref{tab:obs-log}.

\section{Data Reduction}
\label{sec.data-red}

The data reduction used in this paper follows standard imaging methodologies; however, as uGMRT data reduction is a relatively new process, we summarize below the procedure. 
\begin{itemize}
\item[(i)] Periodic observations of calibration source 3C\,286, once every 30 minutes at the 250--500 MHz band and every 20 minutes at the 1050--1450 MHz band were used to correct for the flux density scale and bandpass shape and to perform phase calibration.  The data analysis, including editing of bad data, and gain and bandpass calibrations were carried out using  ``classic" \textsc{aips}.
\item[(ii)] We used the default, revised flux density scale for low frequencies using the coefficients for 3C\,286 in the \textsc{aips} task \textsc{setjy}
$$
\begin{aligned}
{\rm log}(S) = 1.2481 -~0.4507 \times {\rm log}(\nu_G) \\
-~0.1798 \times [{\rm log}(\nu_G)]^2 \\
+~0.0357  \times [{\rm log}(\nu_G)]^3;
\end{aligned}
$$
where $S$ is the spectral flux density and $\nu_G$ is the frequency in GHz \citep[see also][]{PerleyButler}.
\item[(iii)] After the initial calibration, steps (i) and (ii), the 200 MHz of 300--500 MHz data and the 400 MHz of 1050--1450 MHz data were split into five 40~MHz and eight 50~MHz sub-bands, respectively.
These sub-bands were analyzed separately with no averaging of the spectral channels.
\item[(iv)] Areas of $\approx$\,9 deg$^2$ and $\approx$\,1 deg$^2$ were imaged for 300--500 MHz data and 1050--1450 MHz data, respectively; they are just bigger than the first null of the primary beam to image sources far from the phase center and to correct for antenna based gains.

\begin{figure*}[ht]
\begin{center}
\begin{tabular}{c}
\includegraphics[height=16.8cm]{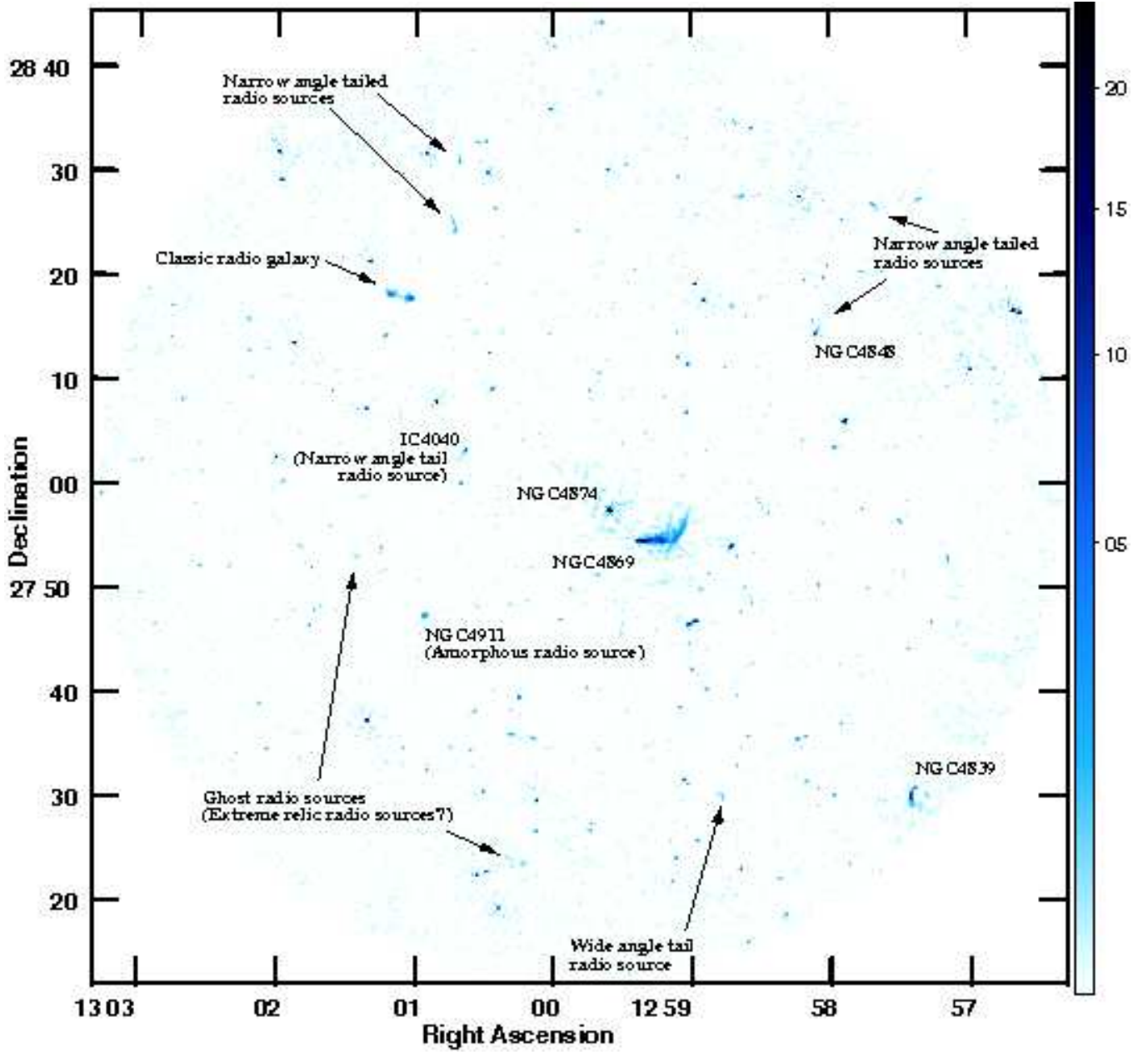}
\end{tabular}
\caption{Image of the Coma cluster of galaxies at the 250--500 MHz band of the uGMRT.
The phase center is RA = 12$^{\rm h}$59$^{\rm m}$ 29$^{\rm s}$ and decl. = 27$^{\rm d}$55$^{\rm m}$48$^{\rm s}$ and the radius (or the field of view) is $\sim$45$^{\prime}$, corresponding to the center frequency, 400 MHz.
The synthesized beam is 6\farcs65 $\times$ 5\farcs90 at a P.A. of 76\fdg39, the peak surface brightness is 114.5 mJy~beam$^{-1}$ and the \textsc{rms} noise is 21 $\mu$Jy~beam$^{-1}$ (see also Sec~\ref{sec.data-red} for a discussion).
The vertical gray scale bar is in units of mJy~beam$^{-1}$.  Representative members of the cluster, including BCG NGC\,4874 along with representative radio sources, candidate extreme relic radio sources, etc., are labeled, showing a variety in radio morphologies.}
\label{ugmrt-halo}
\end{center}
\end{figure*}

\item[(v)] A standard self-calibration procedure and the method of peeling\footnote{First, determine the gains for brightest 5--10 radio sources over the imaged region and subtract these gains from the observed data.
Next, image the residual visibilities after flagging whenever needed.
These two steps were repeated 10--20 times depending on the frequency of the sub-band, which provided us with the gains for $\sim$ 100 directions over the imaged region.  In addition to these brightest ($S_\nu \gtrsim$ 30~mJy) and dominant 32 sources discussed here, we only determined the gains in the directions of strong ($S_\nu \gtrsim$ 10~mJy) compact sources.  We do not perform these steps for the complex/diffuse weak sources, where too many degrees of freedom might add to the problem, and avoid the problem of overfitting \citep{Bhatnagaretal}.} were performed in \textsc{aips} on each sub-band of 300--500 MHz data to correct for direction-dependent errors.
We aimed to obtain the best possible model of all detected sources in the imaged field.
Therefore, we modeled $\approx$100, both extended and pointlike, radio sources that are (nearly) uniformly distributed, providing us with the gains for each direction in the sky at this band.
This peeling step was useful in removing the direction-dependent gains, including the calibration of ionospheric effects, and hence was employed only for 300--500 MHz data; however, only self-calibration procedure was performed for 1050--1450 MHz data.
\item[(vi)] The calibrated data for five 40~MHz and eight 50~MHz sub-bands at the two uGMRT bands were further joined to form full 200 and 400~MHz calibrated visibility data.  The combined visibilities were imaged using \textsc{tclean}.
In order to ensure image fidelity over the full field and band,
we used 3D imaging (gridder = `widefield'), two Taylor coefficients (nterms = 2), and Briggs weighting (robust = 0.5) in task \textsc{tclean}.
\item[(vii)] Using the model from (vi), a final amplitude and phase self-calibration with a solution interval equal to length of observation was then carried out using \textsc{gaincal} and \textsc{applycal} tasks in \textsc{casa}.
The two polarizations, RR and LL were combined to obtain final total intensity image using \textsc{tclean} task in \textsc{casa} as detailed above in step (vi).
We built a task \textsc{wbpbugmrt} using the coefficients of an eighth order polynomial fit to the antenna primary beam for the different bands of the uGMRT\footnote{uGMRT primary beams: \\ \url{www.ncra.tifr.res.in/ncra/gmrt/gmrt-users/observing-help/ugmrt-primary-beam-shape}}, similar to \textsc{widebandpbcor} task in \textsc{casa}.  It computes Taylor-coefficient images that represent the primary beam shape of the GMRT antennas and applies them to the output images from \textsc{tclean} task to produce primary beam corrected images.
\end{itemize}

\begin{figure*}[ht]
\begin{center}
\begin{tabular}{c}
\includegraphics[height=11.0cm]{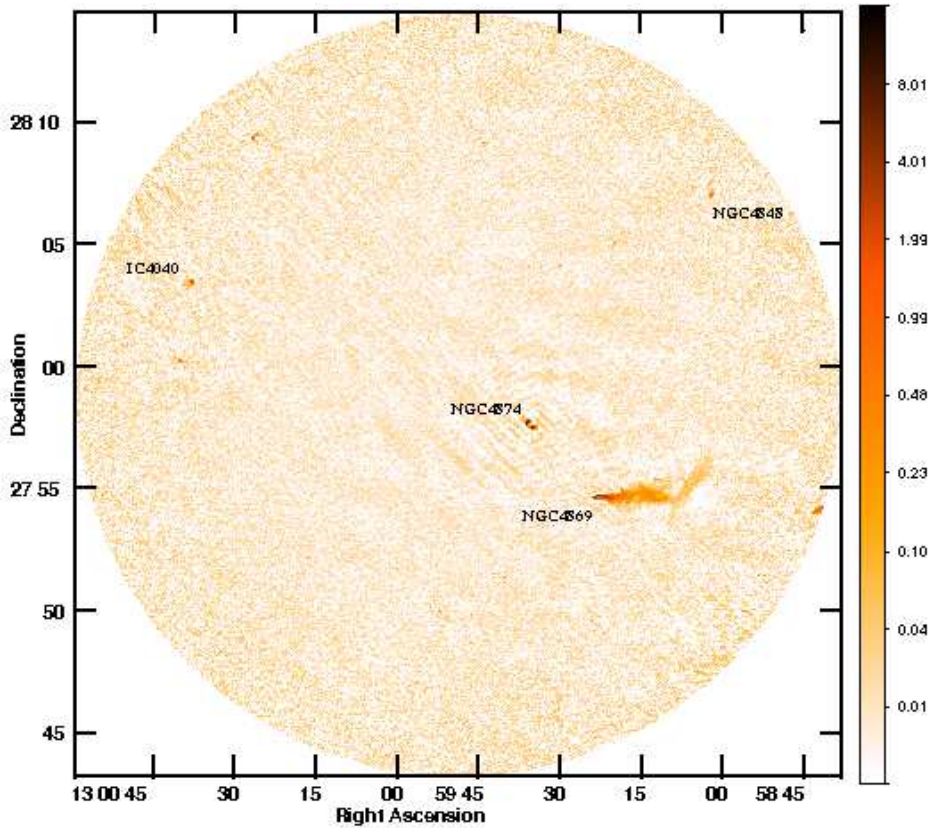}
\end{tabular}
\caption{Image of the Coma cluster of galaxies at the 1050--1450 MHz band of the uGMRT.
The radius (or the field of view) is $\sim$15.6$^{\prime}$, corresponding to the center frequency, 1250 MHz.
The synthesized beam is 2\farcs43 $\times$ 1\farcs95 at a P.A. of 69\fdg51,
the peak surface brightness is 22.5 mJy~beam$^{-1}$ and the \textsc{rms} noise is 12.7 $\mu$Jy~beam$^{-1}$ (see also Sec~\ref{sec.data-red} for a discussion).
The vertical gray scale bar is in units of mJy~beam$^{-1}$.}
\label{ugmrt-lband}
\end{center}
\end{figure*}

The final images corrected for the primary beam shape
of the GMRT antennas for the two bands, shown in Fig.~\ref{ugmrt-halo} and Fig.~\ref{ugmrt-lband}, have 
an \textsc{rms} noise of $\approx$ 21~$\mu$Jy~beam$^{-1}$ and $\approx$ 13~$\mu$Jy~beam$^{-1}$ at the half power point and a dynamic range of $\approx$~5300 and $\approx$~1700, respectively.
The error in the estimated flux density, both due to calibration and due to systematic, is $\lesssim$4\%.
The \textsc{rms} noise is a factor of 1.8--2.1 higher close to the phase center where two dominant radio sources and an extended radio source, NGC\,4874, the BCG and NGC\,4869, are present.
Subsequent analyses of the integrated spectra of all these sources shows that our flux densities measurements are consistent with the literature data (see also Sec.~\ref{rad-spec}).

\section{Results}
\label{sec.morph-spec}

The large scale radio morphology of the Coma cluster at the 250--500 MHz band and the 1050--1450 MHz band of the uGMRT is shown in Fig.~\ref{ugmrt-halo} and Fig.~\ref{ugmrt-lband}, respectively.
These are the deepest GMRT images after its upgrade (at the 250--500 MHz band) to our knowledge.
Fig.~\ref{collage-1} shows images of a subset of the 32 brightest ($S_\nu \gtrsim 30$~mJy) and dominant sources at the 250--500 MHz band and the majority of these show discrete pointlike radio morphologies.
Similarly, Fig.~\ref{collage-Lband} shows images of 5 (out of 32) sources imaged in the $\sim$15\farcm6 field of view at the 1050--1450 MHz band.
The flux density is determined using \textsc{aips} task \textsc{jmfit} for pointlike, symmetric sources and using \textsc{aips} task \textsc{tvstat} for all extended (source sizes $\gtrsim$ 15$^{\prime\prime}$) irregularly shaped radio sources.
The error bars in Table~3 are based on the local \textsc{rms} noise as evaluated in a circle of 5$^{\prime}$ in diameter centered on the source position; this is also the error associated with the peak flux density measurement.
Table~\ref{tab:peel-model} lists these 32 modeled, brightest ($S_\nu \gtrsim$ 30~mJy), and both pointlike and extended radio sources present in the field of view (90$^{\prime}$ $\pm$4$^{\prime}$) at 250--500 MHz band after correcting for the primary beam shape.
The columns are  as  follows: (1) source ID; (2 and 3) likely radio core position (R.A. and decl.) at 400 MHz; (4) redshift of source, from NED; and
(5) comments on the source structure at the 250--500~MHz band.

\subsection{Notes on the dominant radio sources}
\label{notes-dominant}

Salient features of radio morphologies of these brightest ($S_\nu \gtrsim 30$~mJy) and dominant radio sources are presented in Table~\ref{tab:peel-model}.
Below we provide additional notes on our interpretation of the radio morphologies for these radio sources in the order presented in Table~\ref{tab:peel-model}.
We also include notes of the radio morphologies for five sources, Source\_IDs, 01, 02, 09, 17 and 20, that were imaged at the 1050--1450 MHz band. 

\begin{figure*}[ht]
\begin{center}
\begin{tabular}{cccc}
\includegraphics[height=3.93cm]{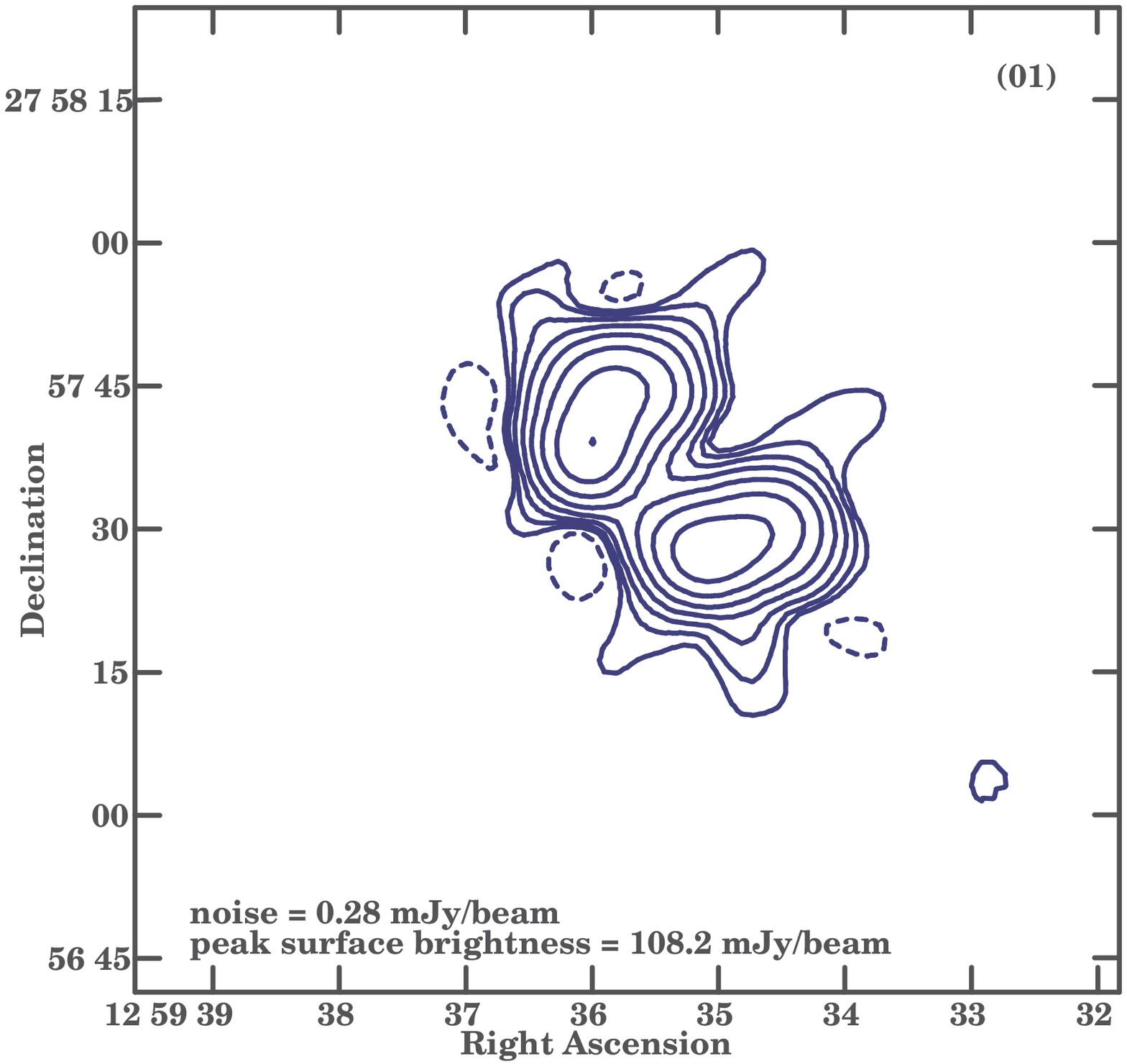} &
\includegraphics[height=3.93cm]{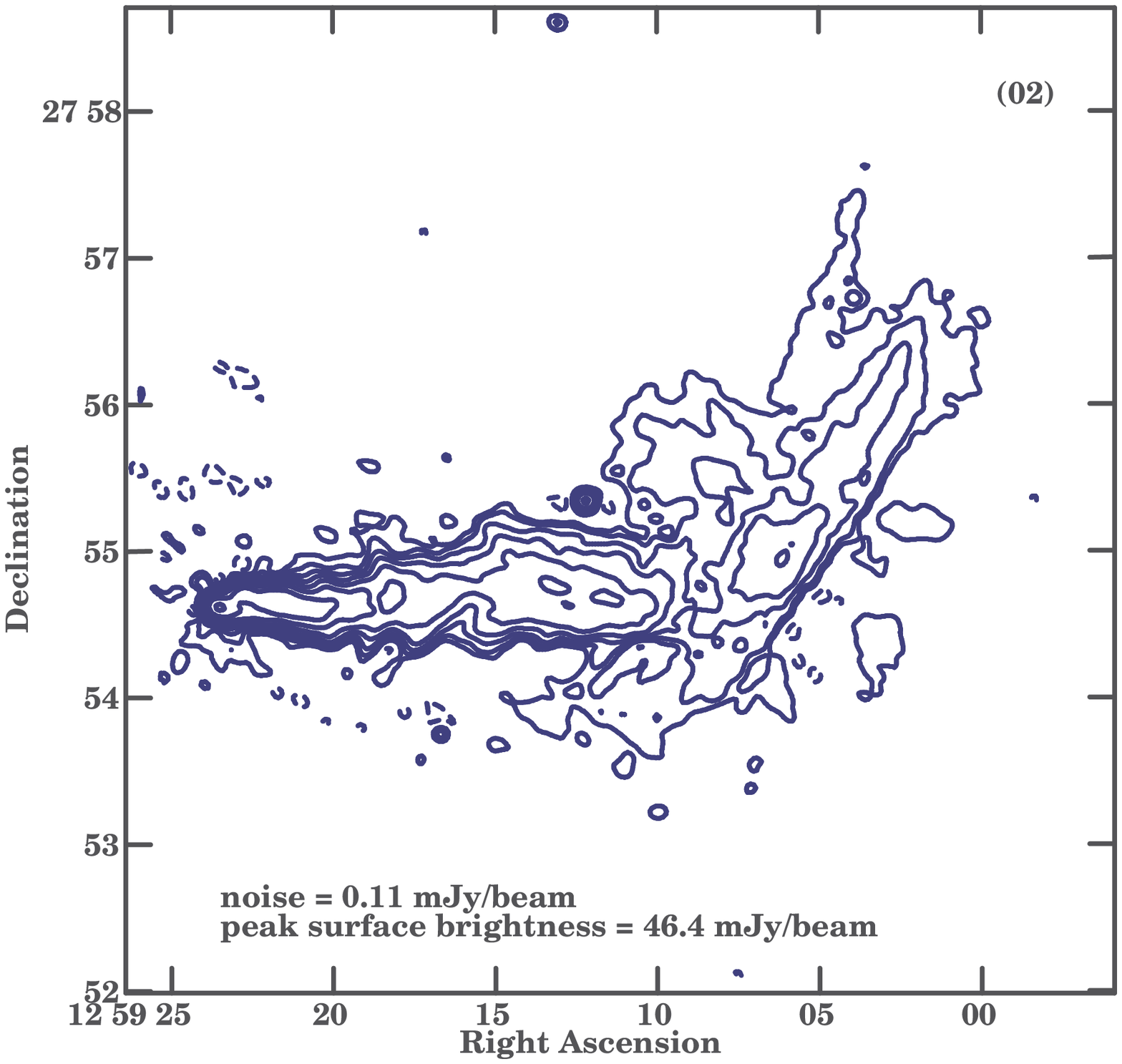} &
\includegraphics[height=3.93cm]{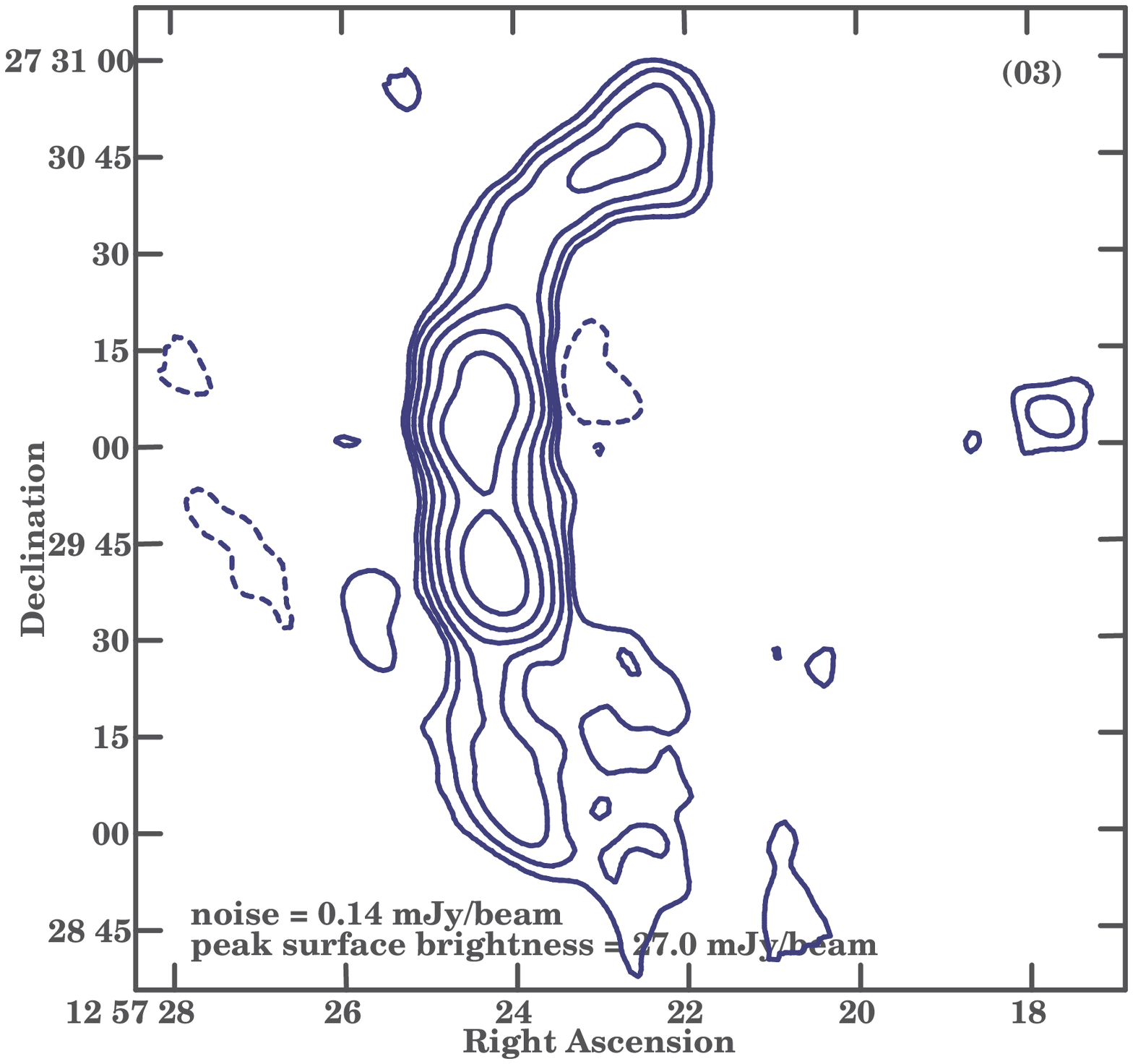} &
\includegraphics[height=3.93cm]{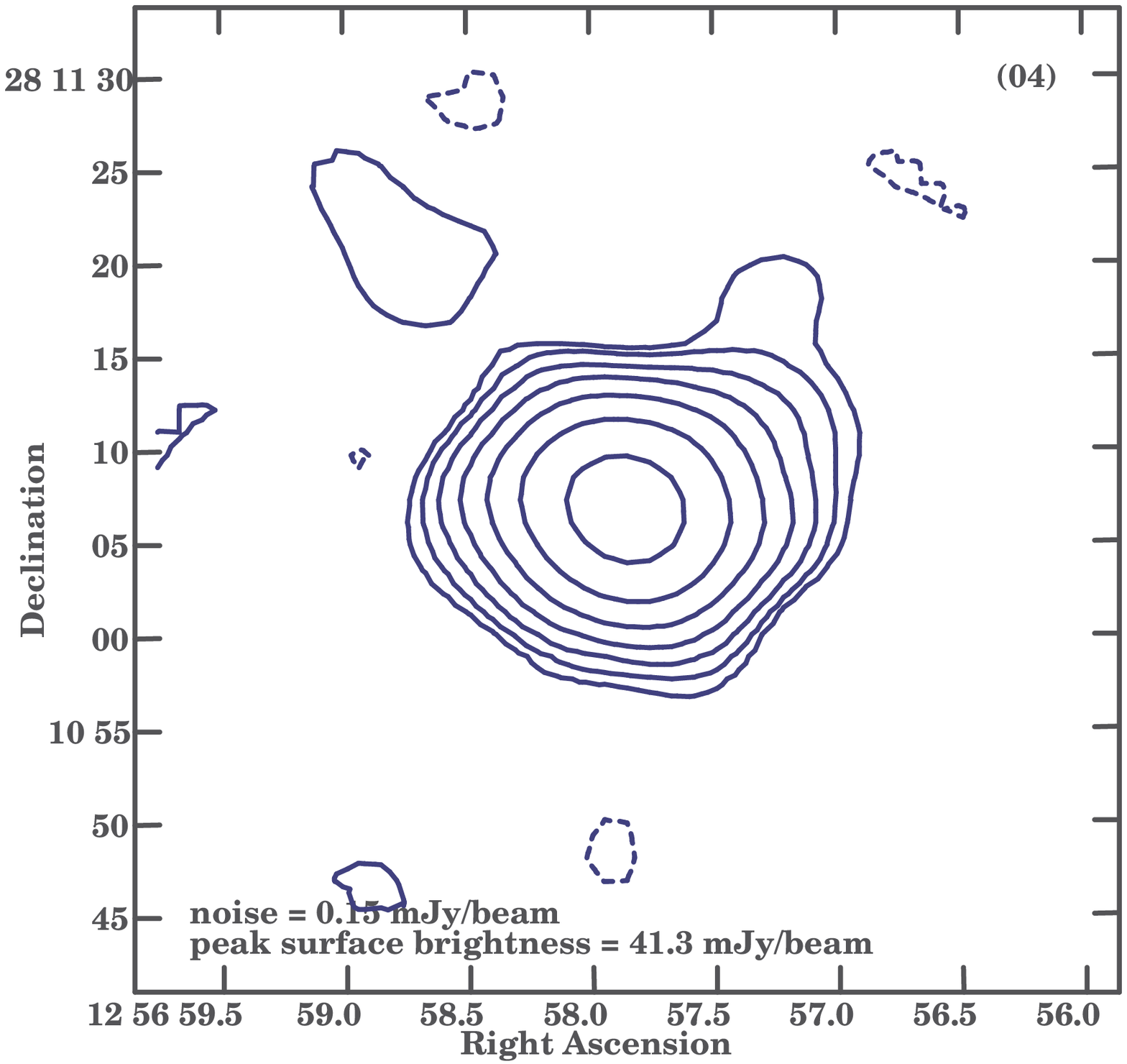} \\
\includegraphics[height=3.93cm]{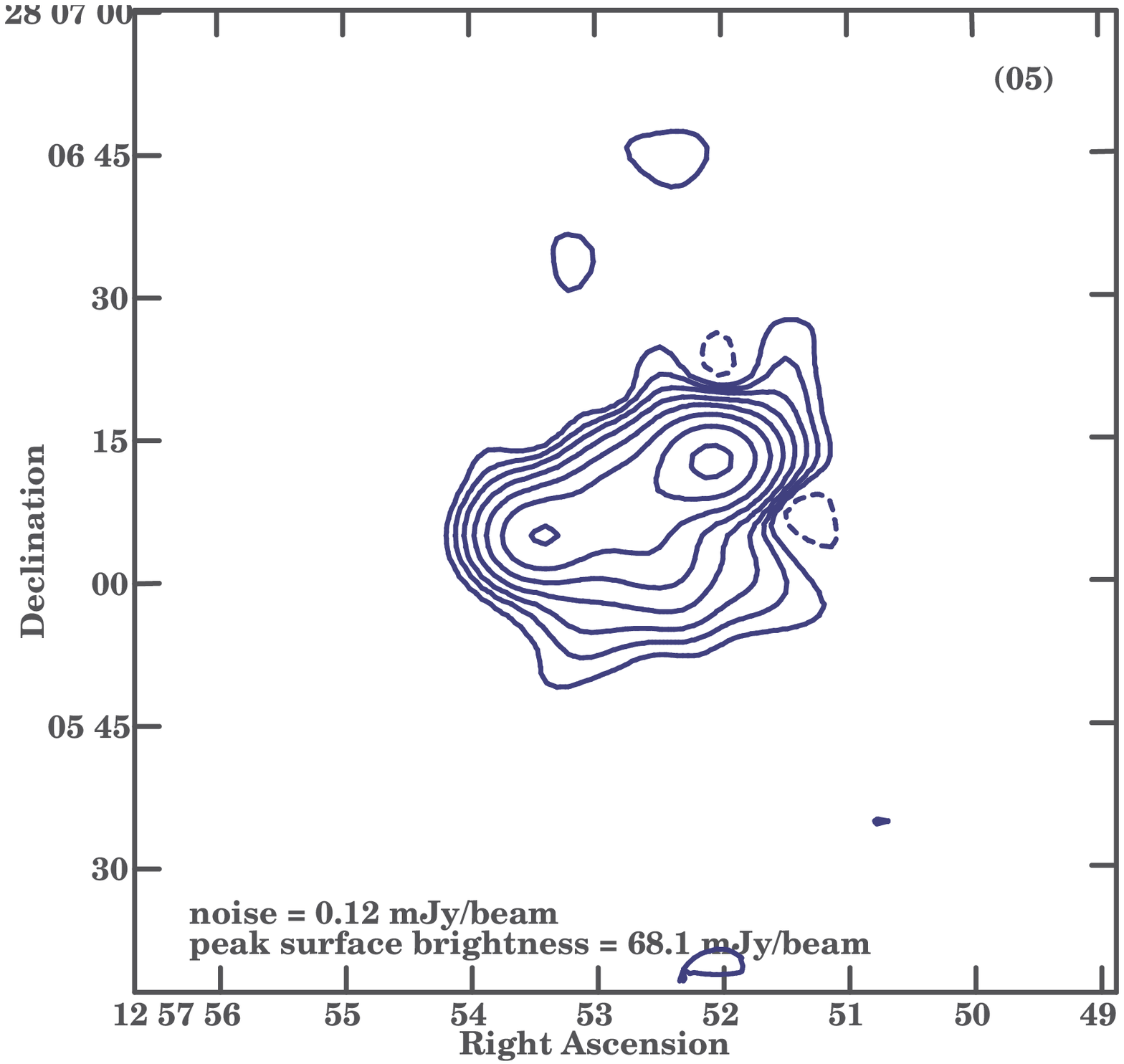} &
\includegraphics[height=3.93cm]{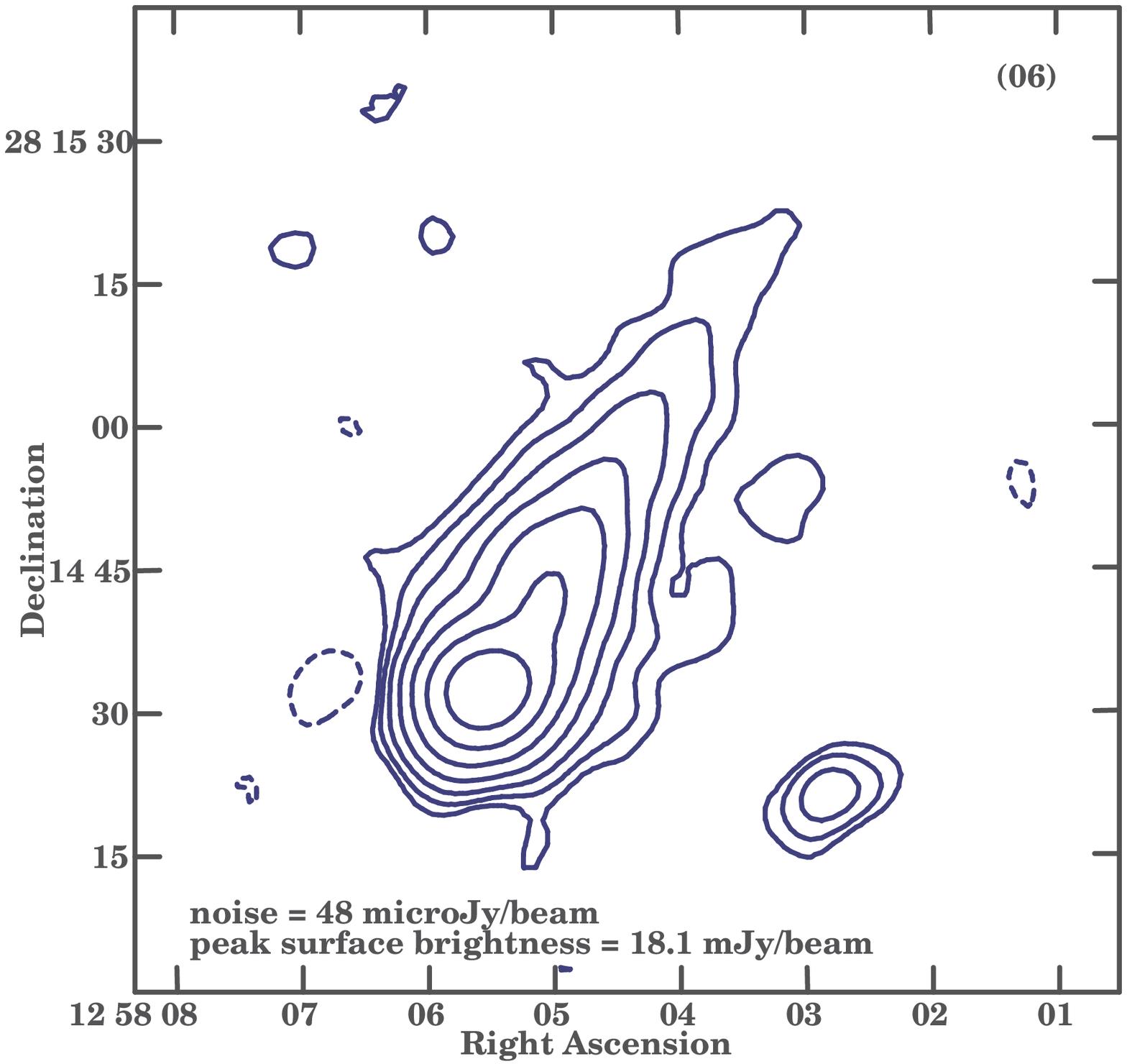} &
\includegraphics[height=3.93cm]{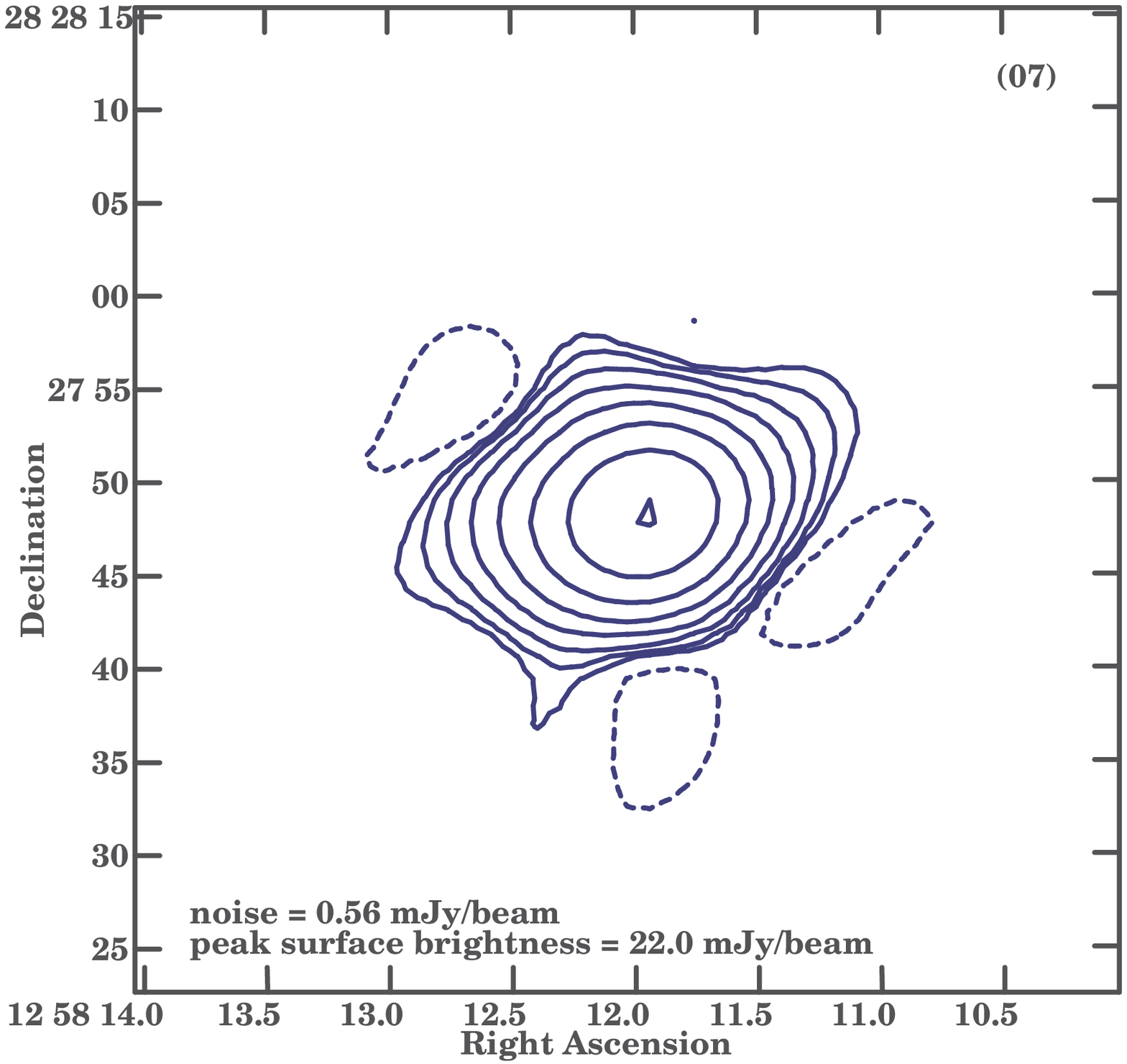} &
\includegraphics[height=3.93cm]{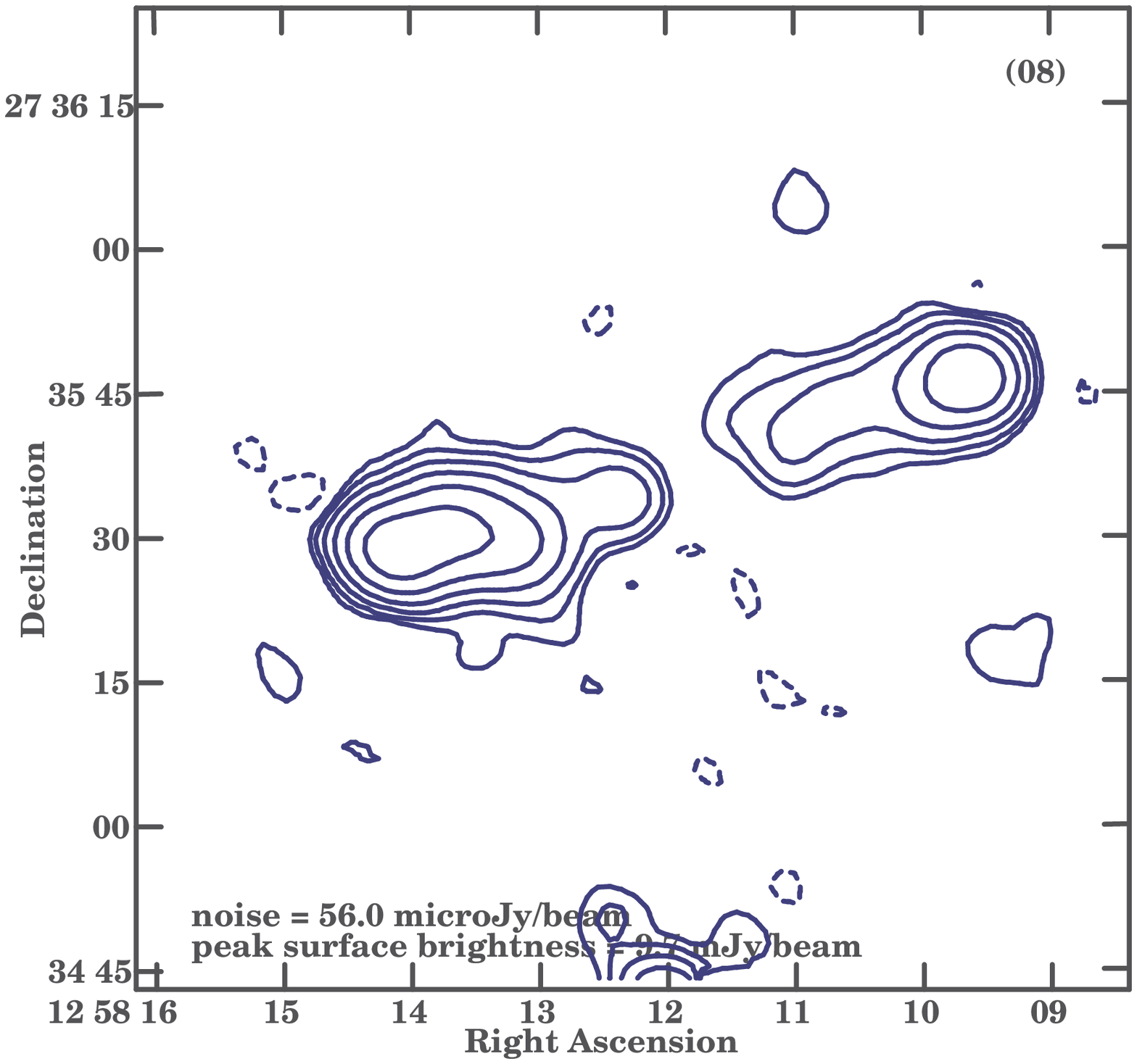} \\
\includegraphics[height=3.93cm]{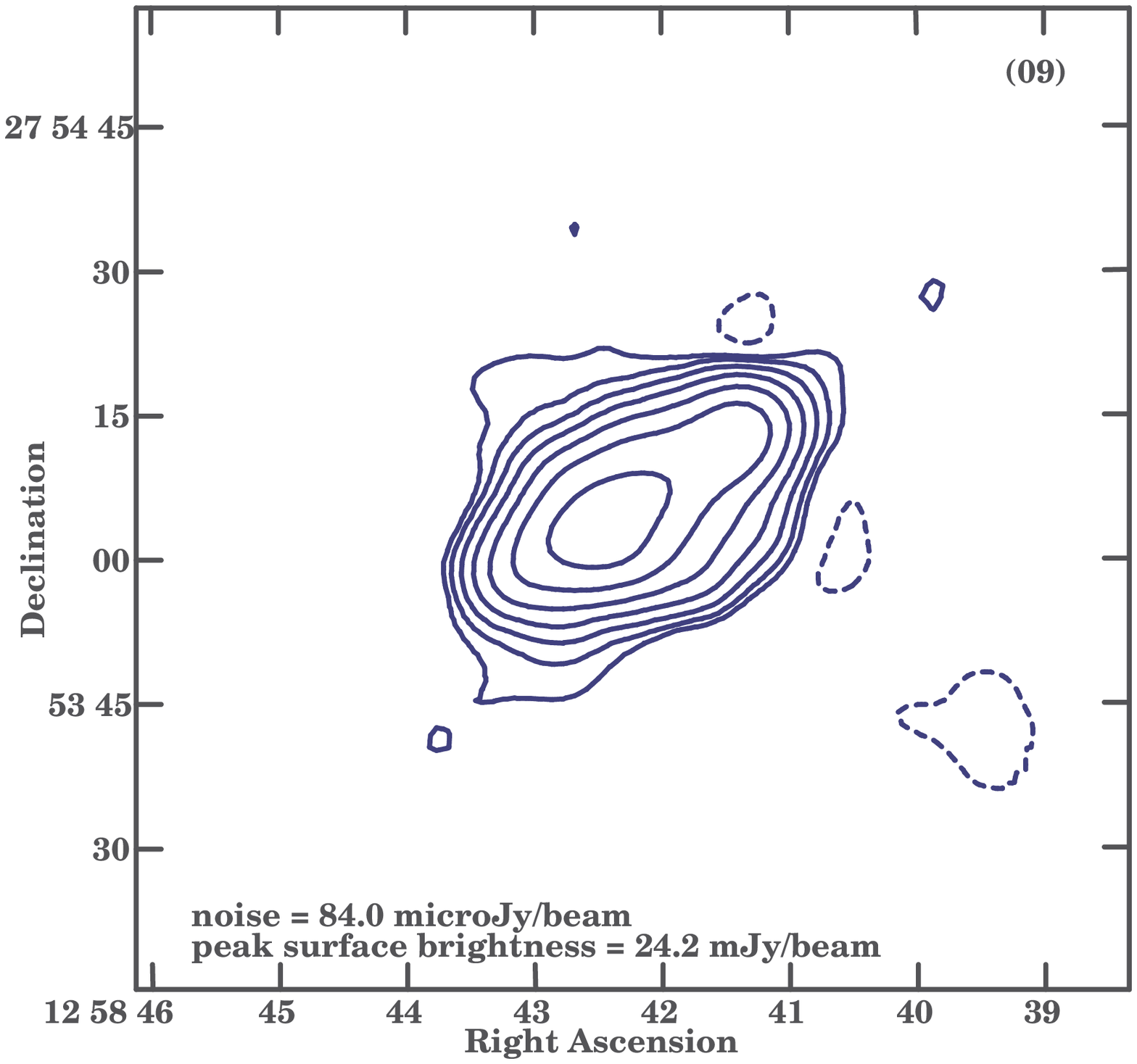} &
\includegraphics[height=3.93cm]{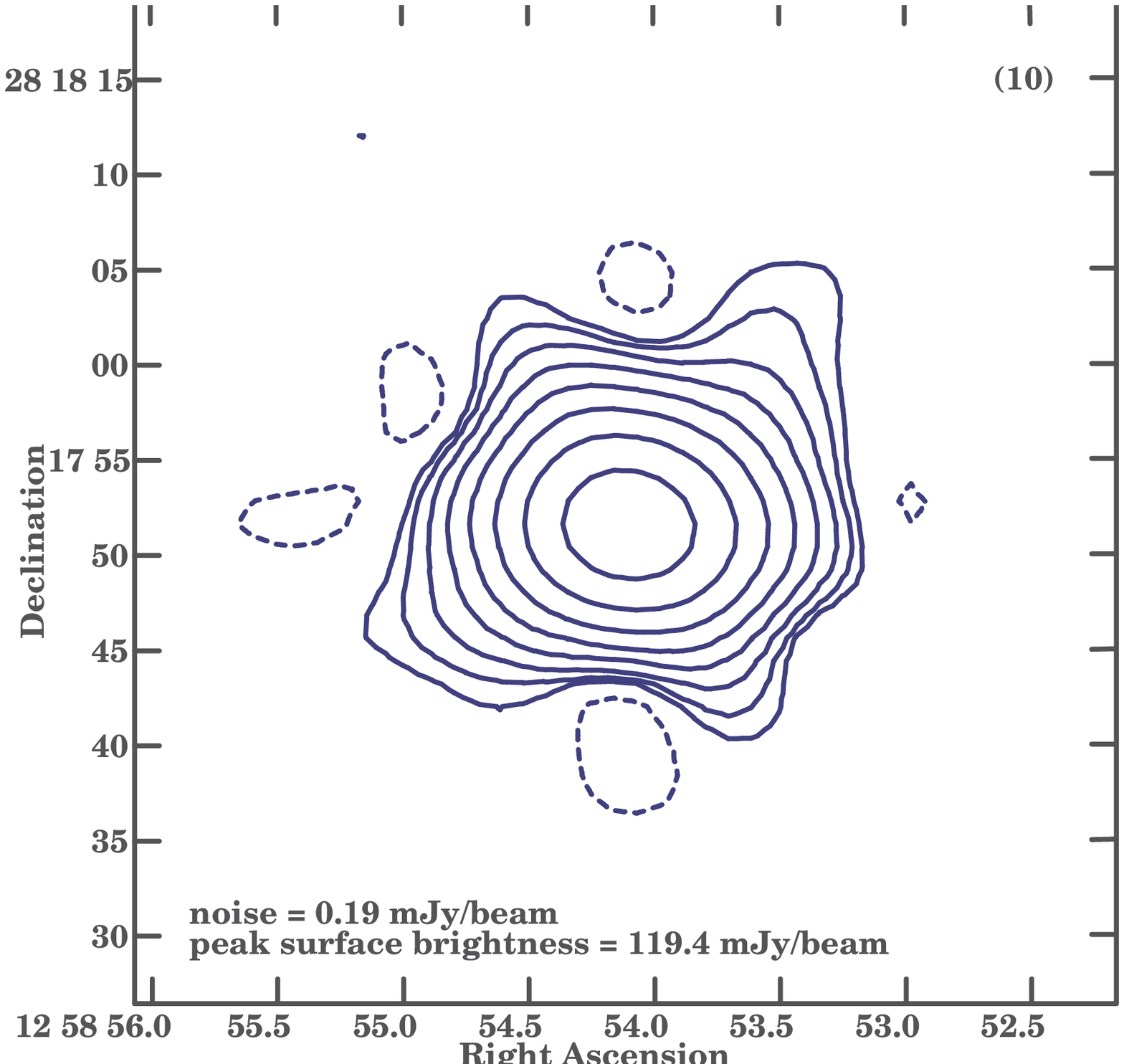} &
\includegraphics[height=3.93cm]{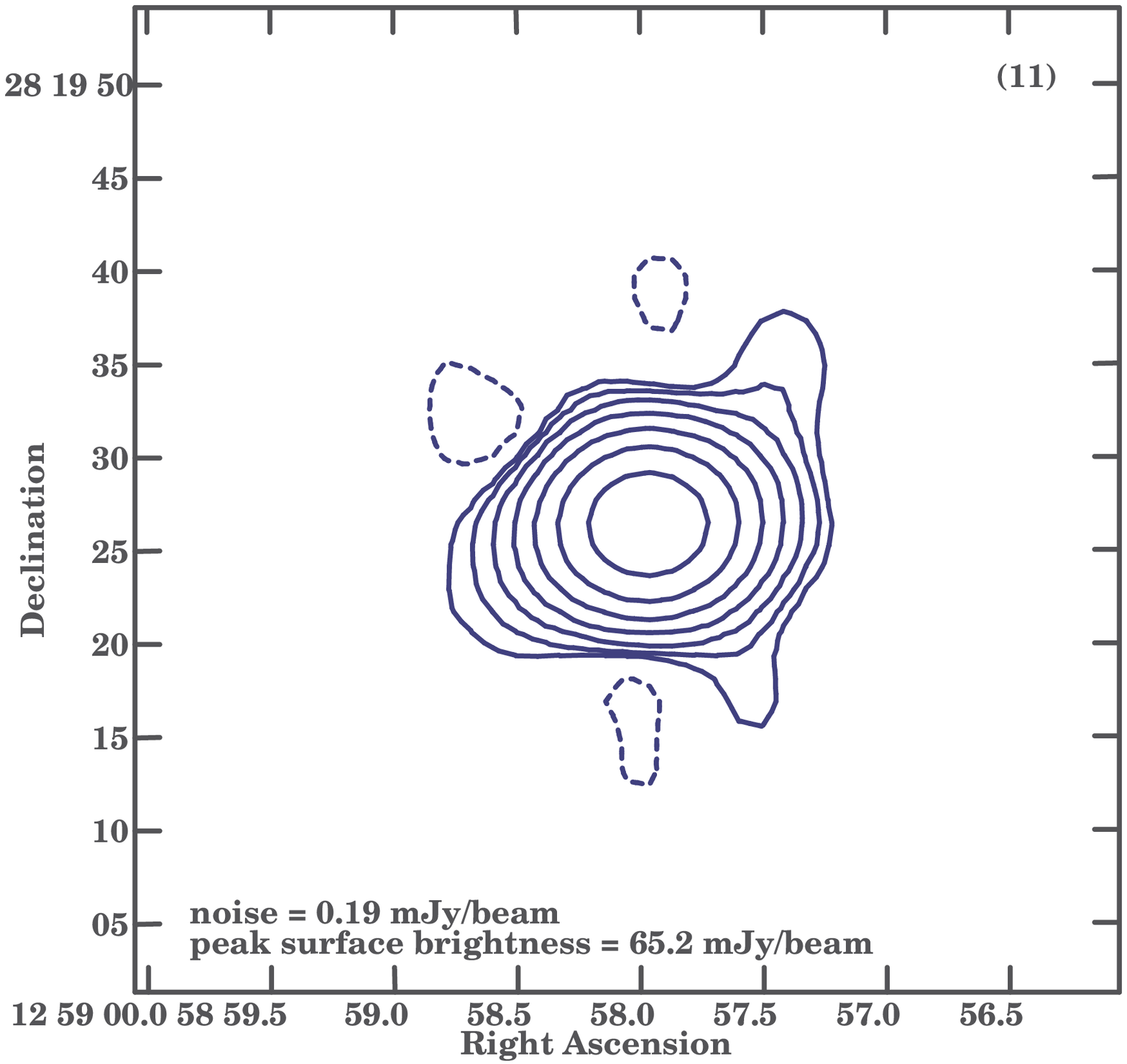} &
\includegraphics[height=3.93cm]{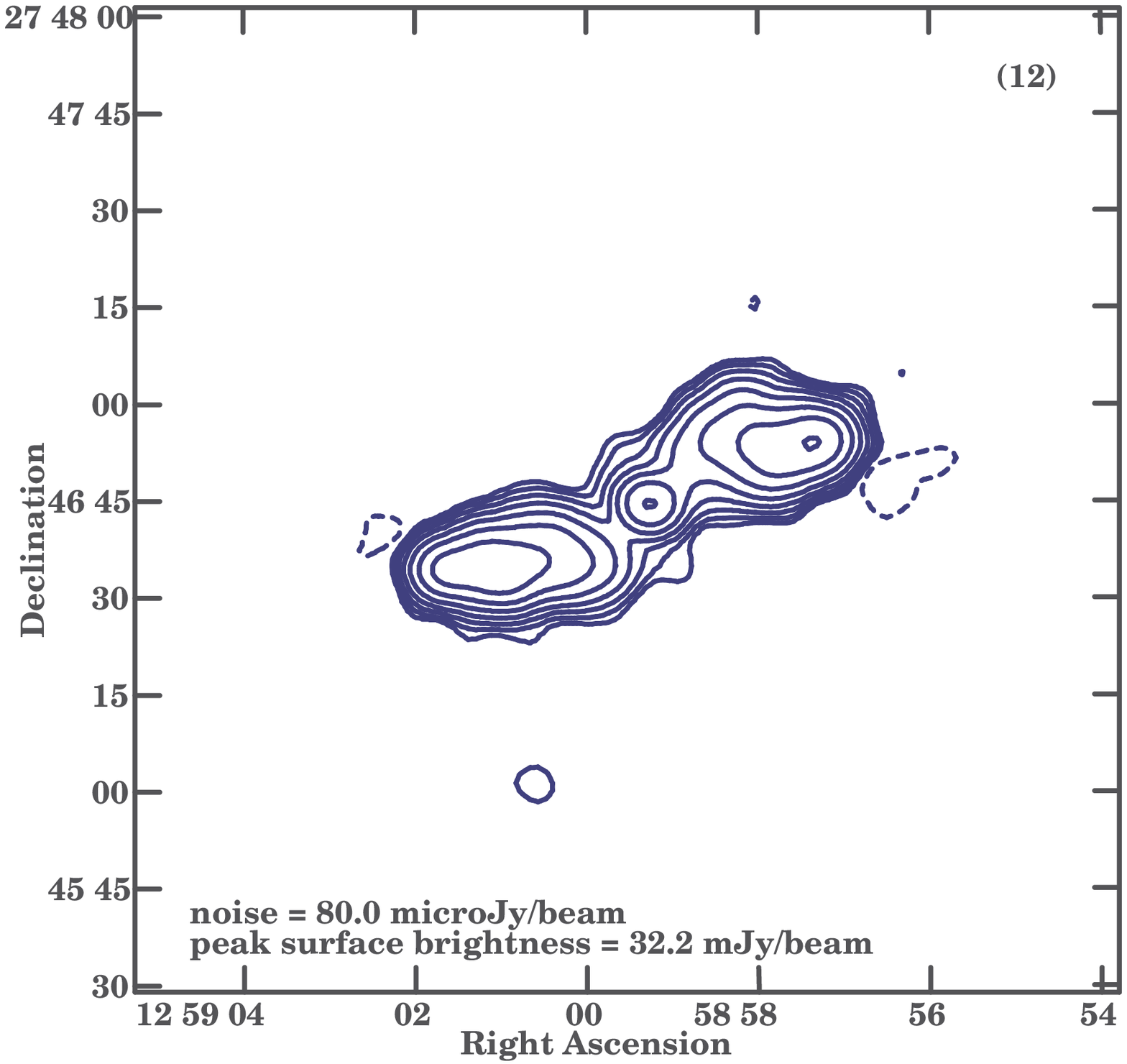} \\
\includegraphics[height=3.93cm]{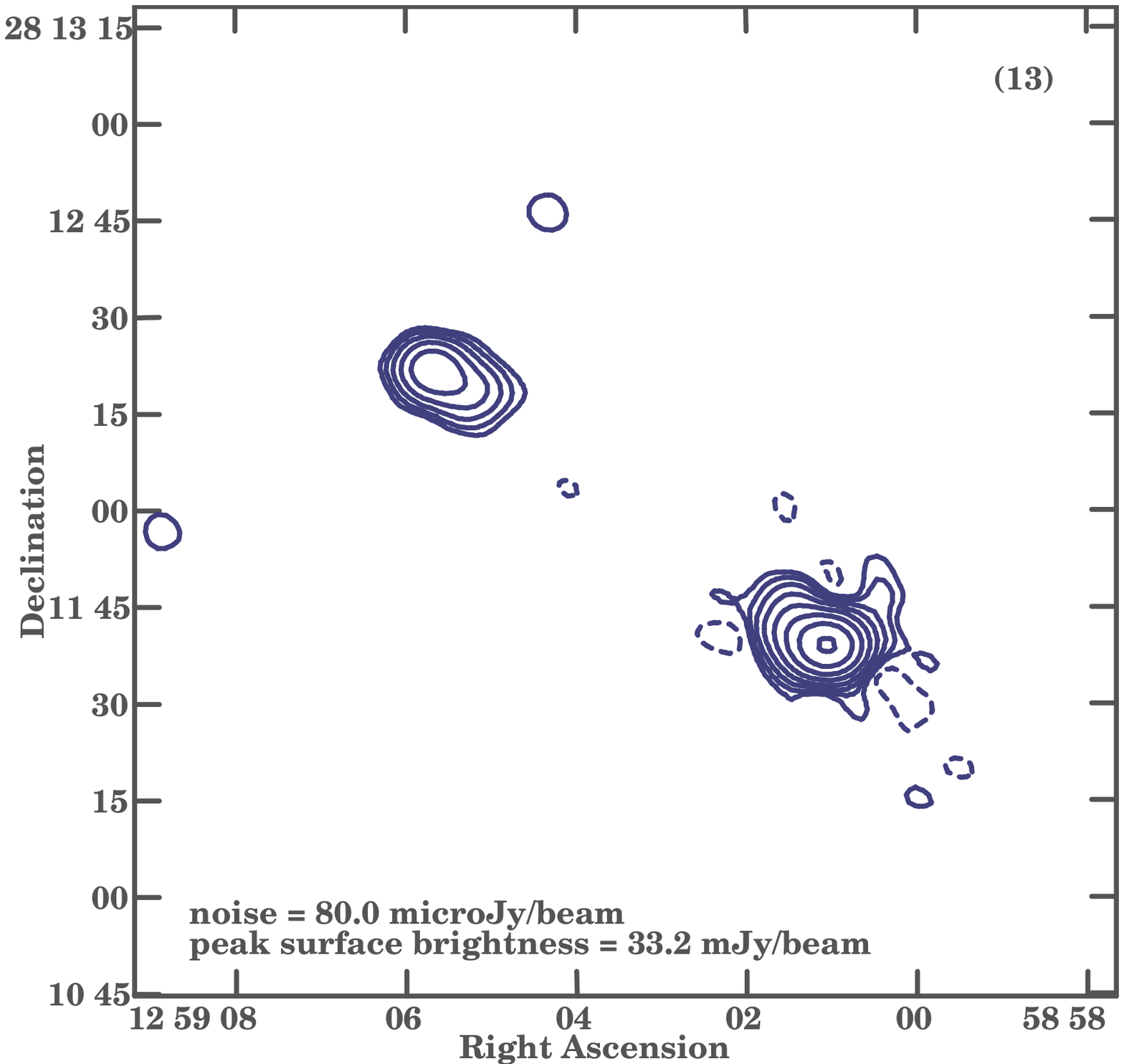} &
\includegraphics[height=3.93cm]{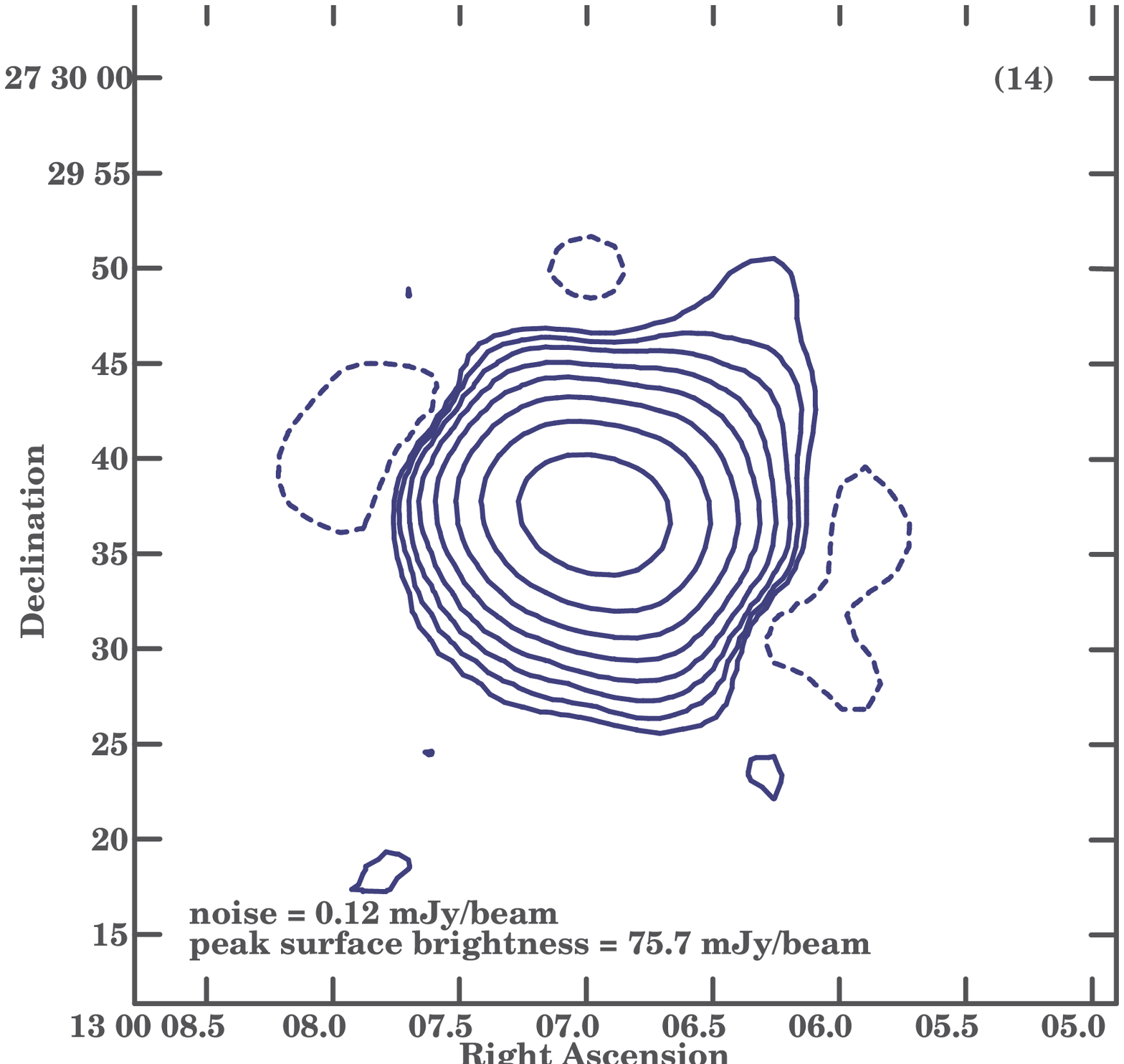} &
\includegraphics[height=3.93cm]{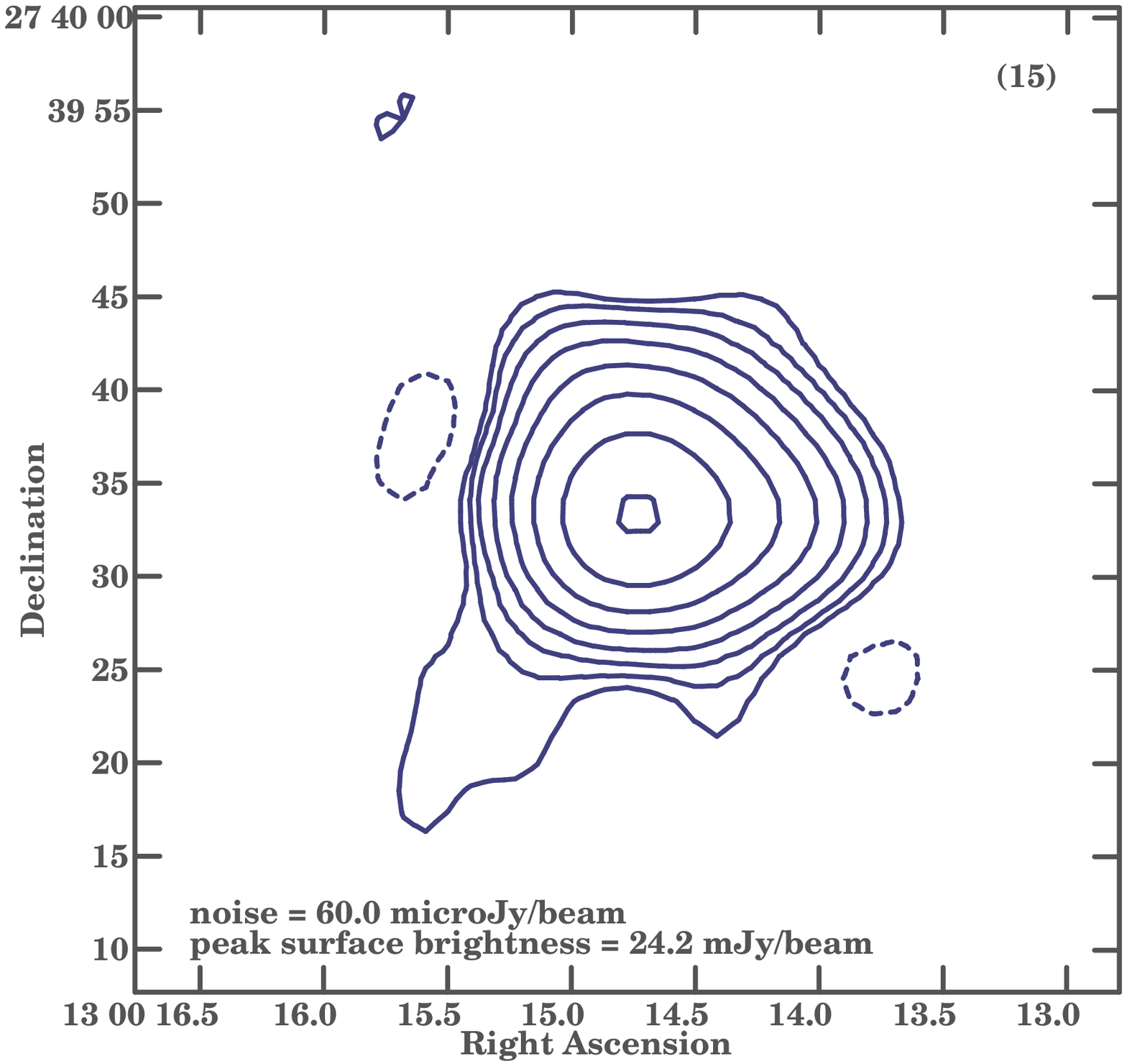} &
\includegraphics[height=3.93cm]{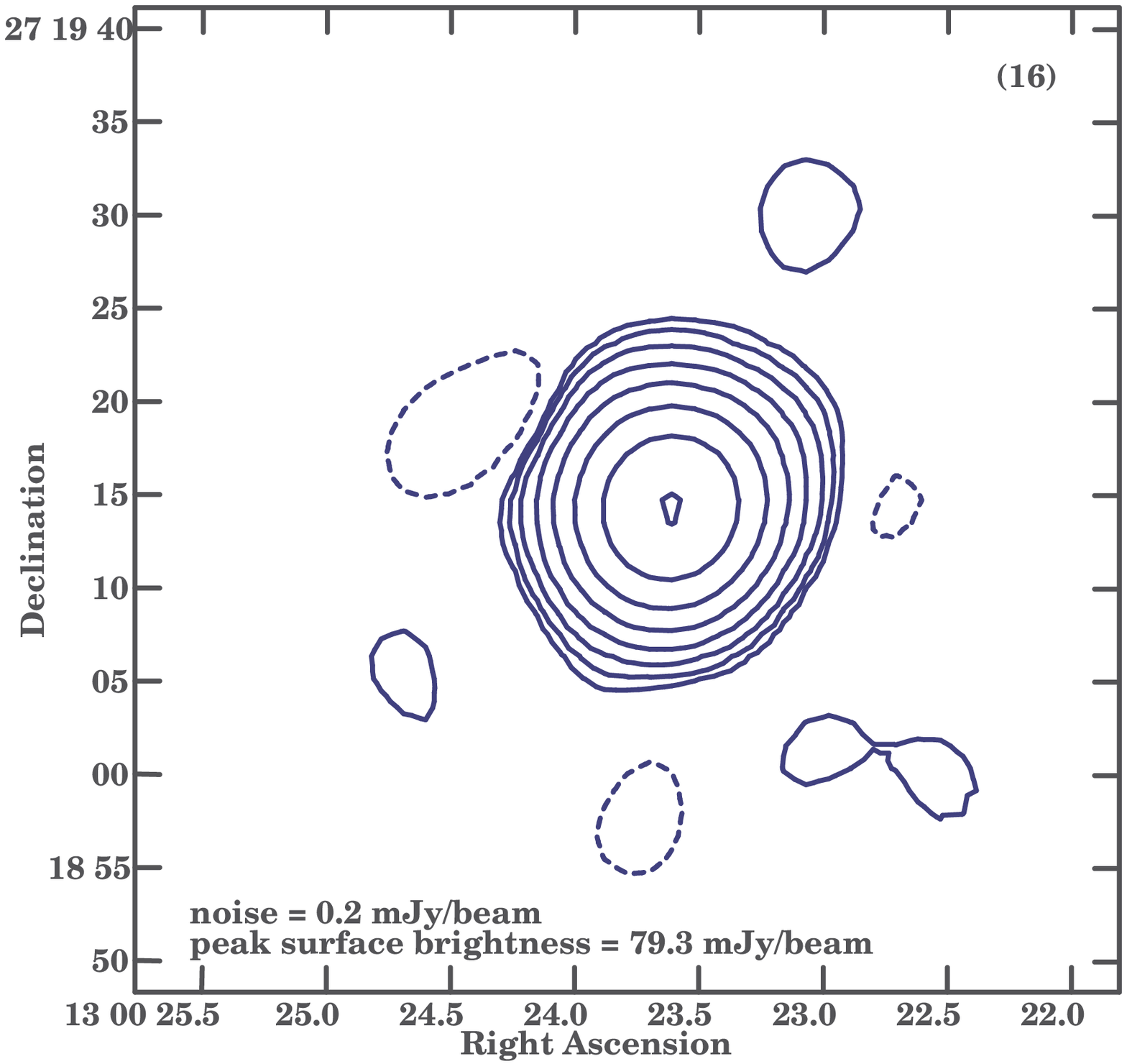}
\end{tabular}
\caption{Images of brightest ($S_\nu \gtrsim$ 30~mJy) and dominant, both pointlike and extended, radio sources in the Coma cluster at 250--500 MHz band of the uGMRT.
These radio sources are in the order presented in Table~\ref{tab:peel-model}.
The lowest radio contour plotted is three times the local \textsc{rms} noise and increasing by factors of 2.  The local \textsc{rms} noise and continuum peak surface brightness of the source are denoted in each panel (lower-left corner) along with its Source\_ID (upper-right corner).}
\label{collage-1}
\end{center}
\end{figure*}

\begin{itemize}
\item[01.] NGC\,4874 $-$
The dominant BCG of the Coma cluster \citep{BaierTiersch1990}
that is located close to the peak of the distributions of galaxies and
X-ray gas.  The optical host is a regular low brightness elliptical galaxy
\citep{2000A&A...362..871C} located at $z$ = 0.02394.
It is also a strong radio source with a wide-angle tail (WAT) radio morphology and its projected maximum angular extension is $\sim$30$^{\prime\prime}$, corresponding to $\sim$15~kpc.
We detect the radio core and two radio jets forming a wide angle between them in our uGMRT 250--500 MHz band image.

The source is a dominant cD radio galaxy, where the local thermal noise is a factor of $\sim$2.1 higher than at the half power point.  In the vicinity of this dominant source, the dynamic range is smallest leading to residual deconvolution (and hence phase calibration) errors ($\lesssim$1\%) at the 1050--1450\,MHz band

\setcounter{figure}{2}
\begin{figure*}[ht]
\begin{center}
\begin{tabular}{cccc}
\includegraphics[height=3.93cm]{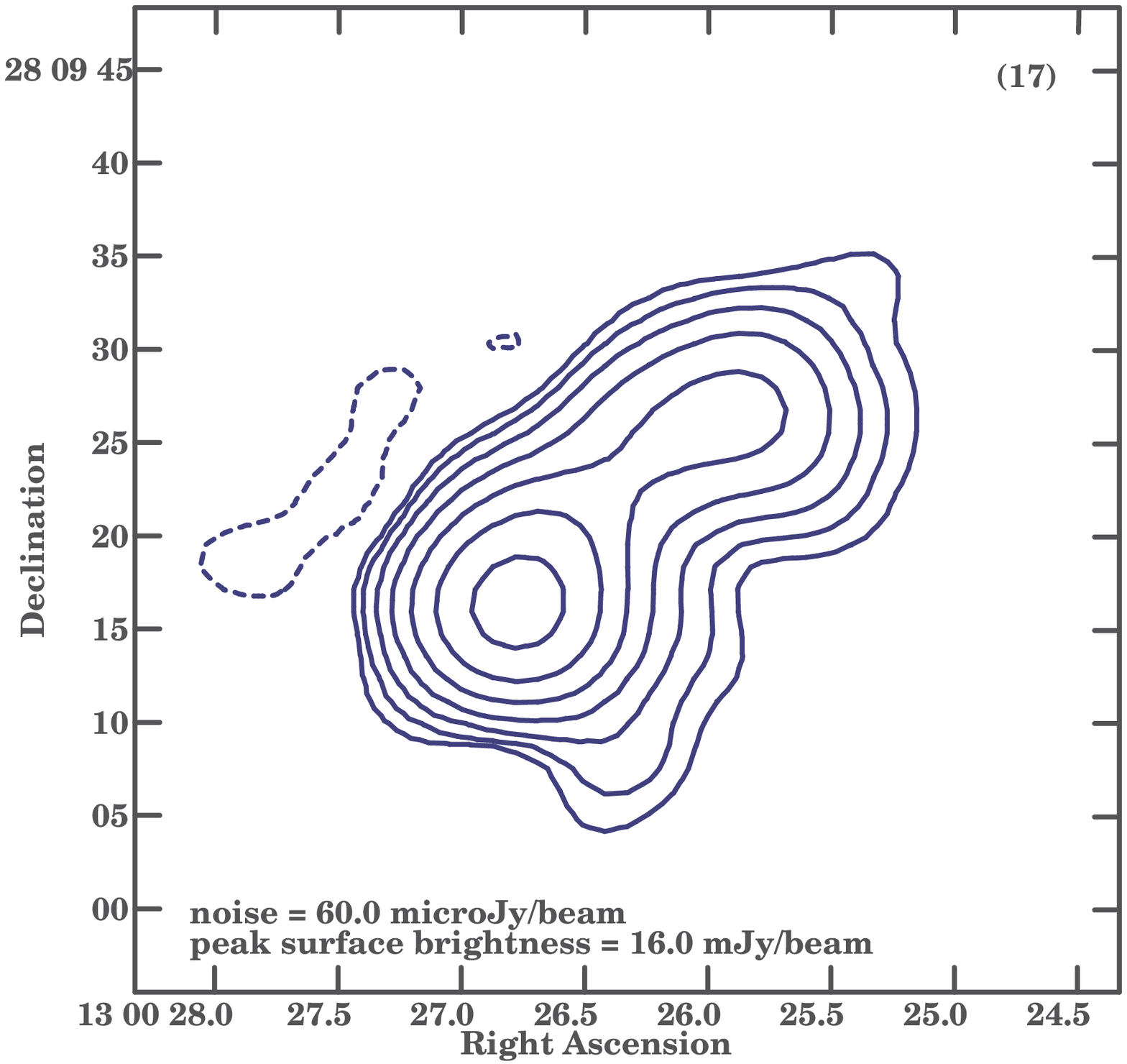} &
\includegraphics[height=3.93cm]{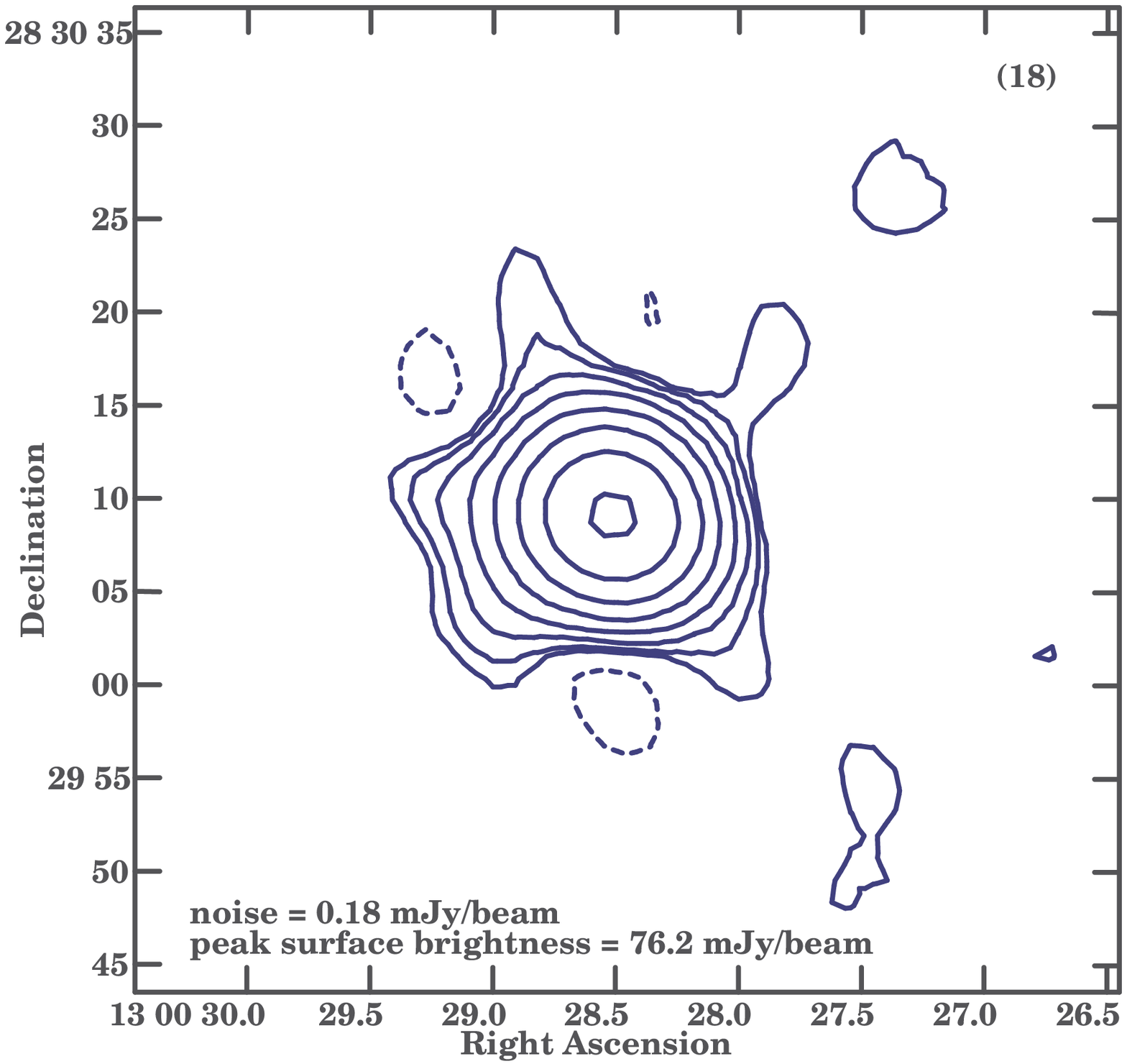} &
\includegraphics[height=3.93cm]{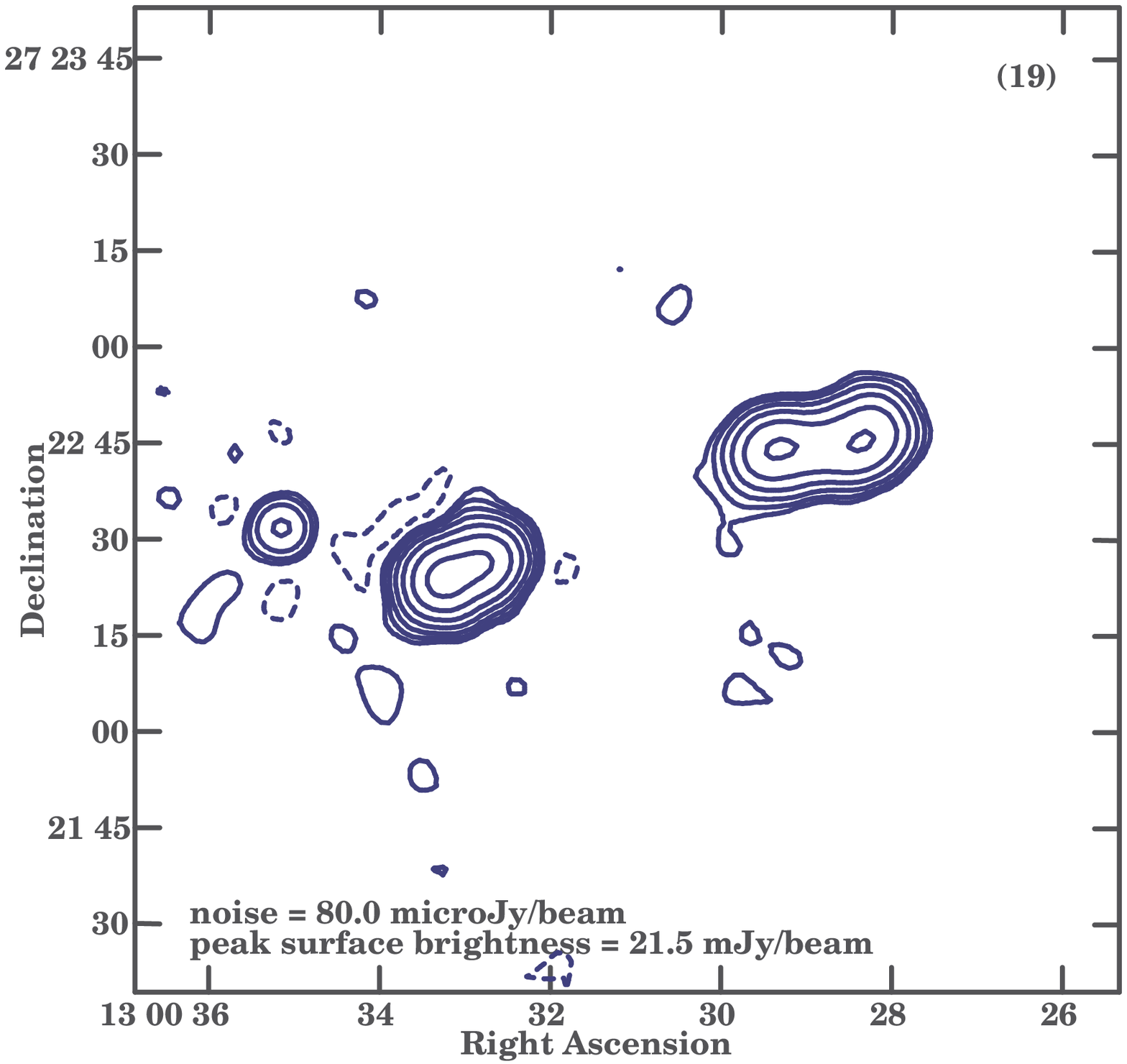} &
\includegraphics[height=3.93cm]{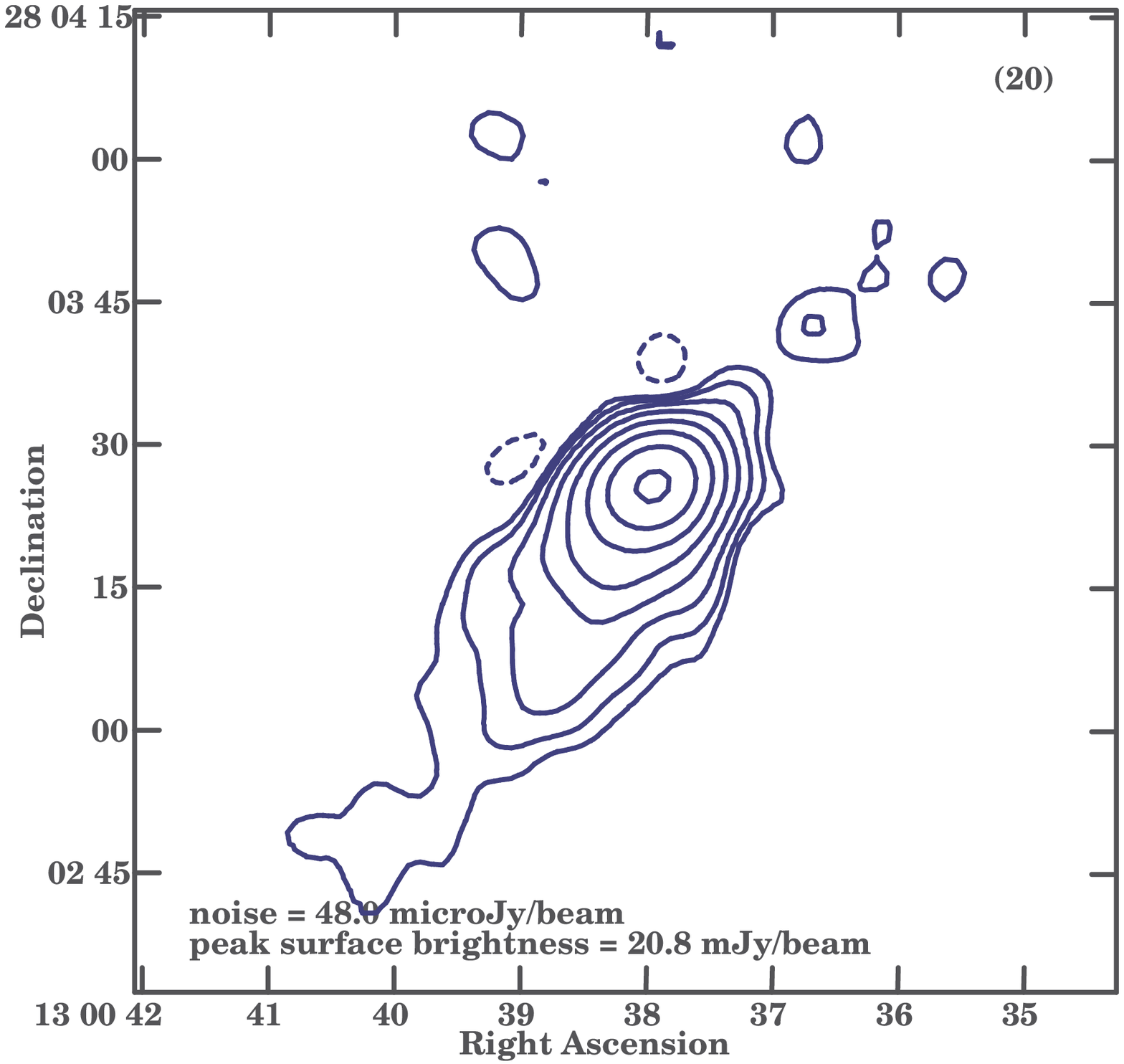} \\
\includegraphics[height=3.93cm]{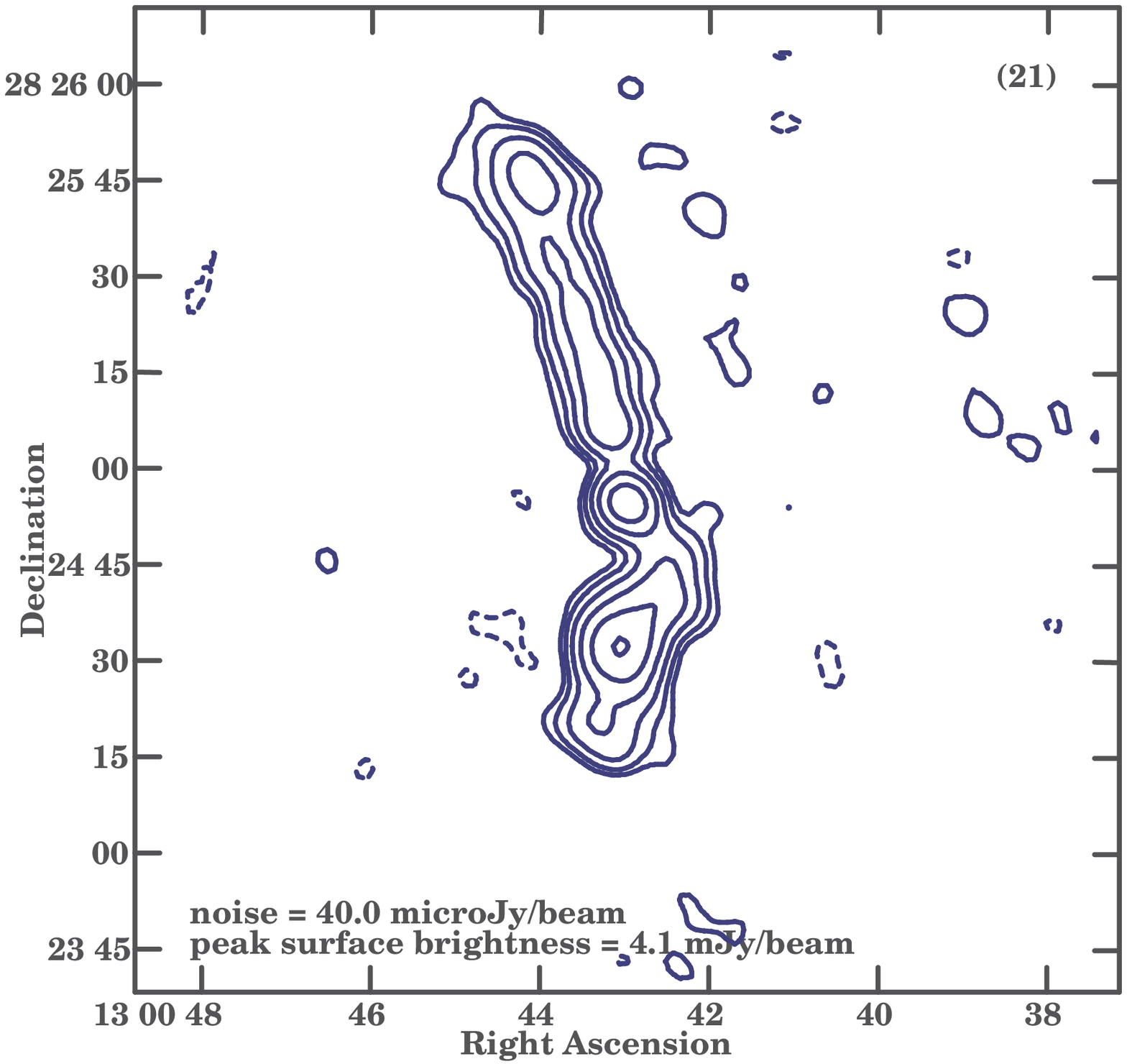} &
\includegraphics[height=3.93cm]{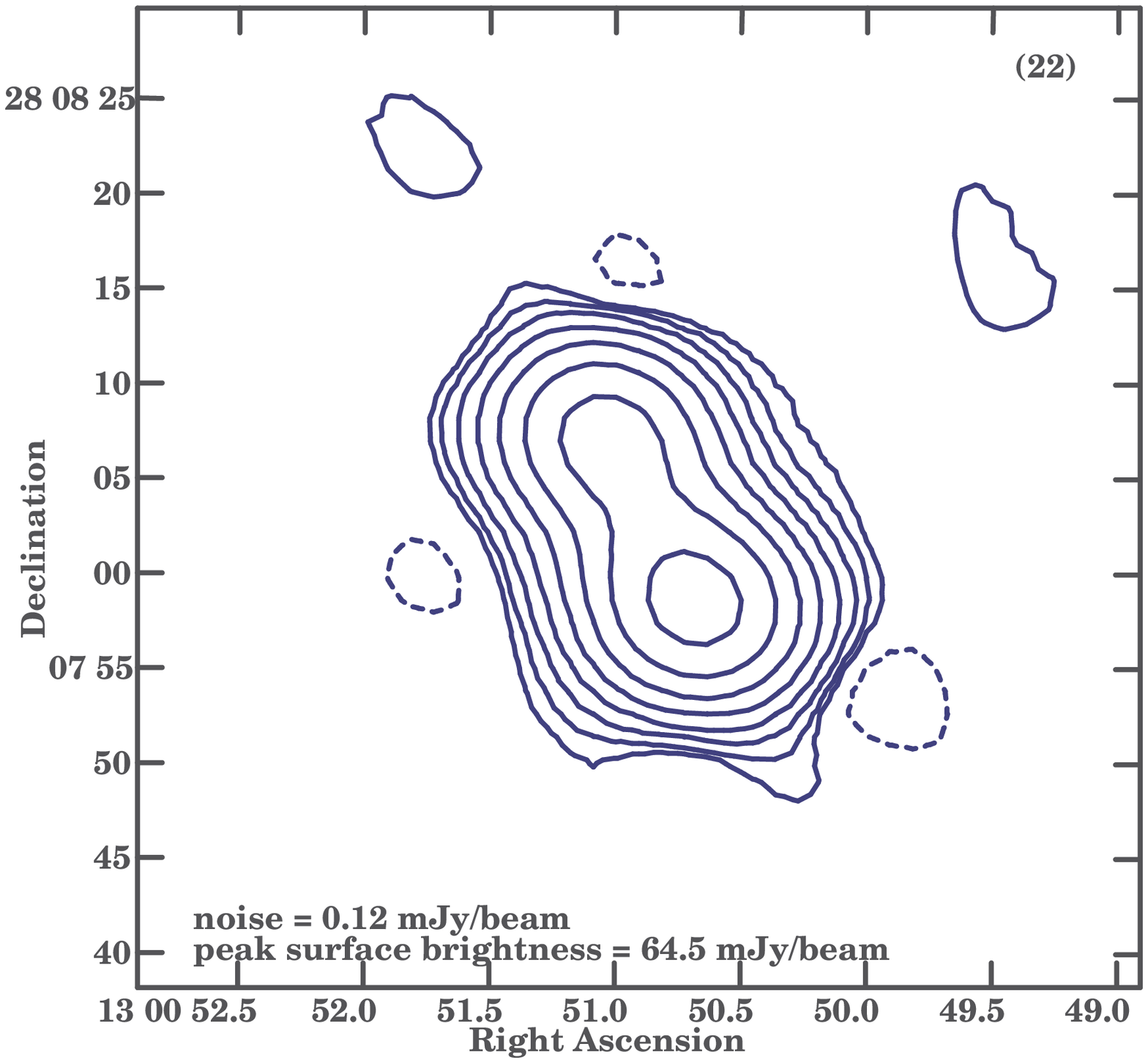} &
\includegraphics[height=3.93cm]{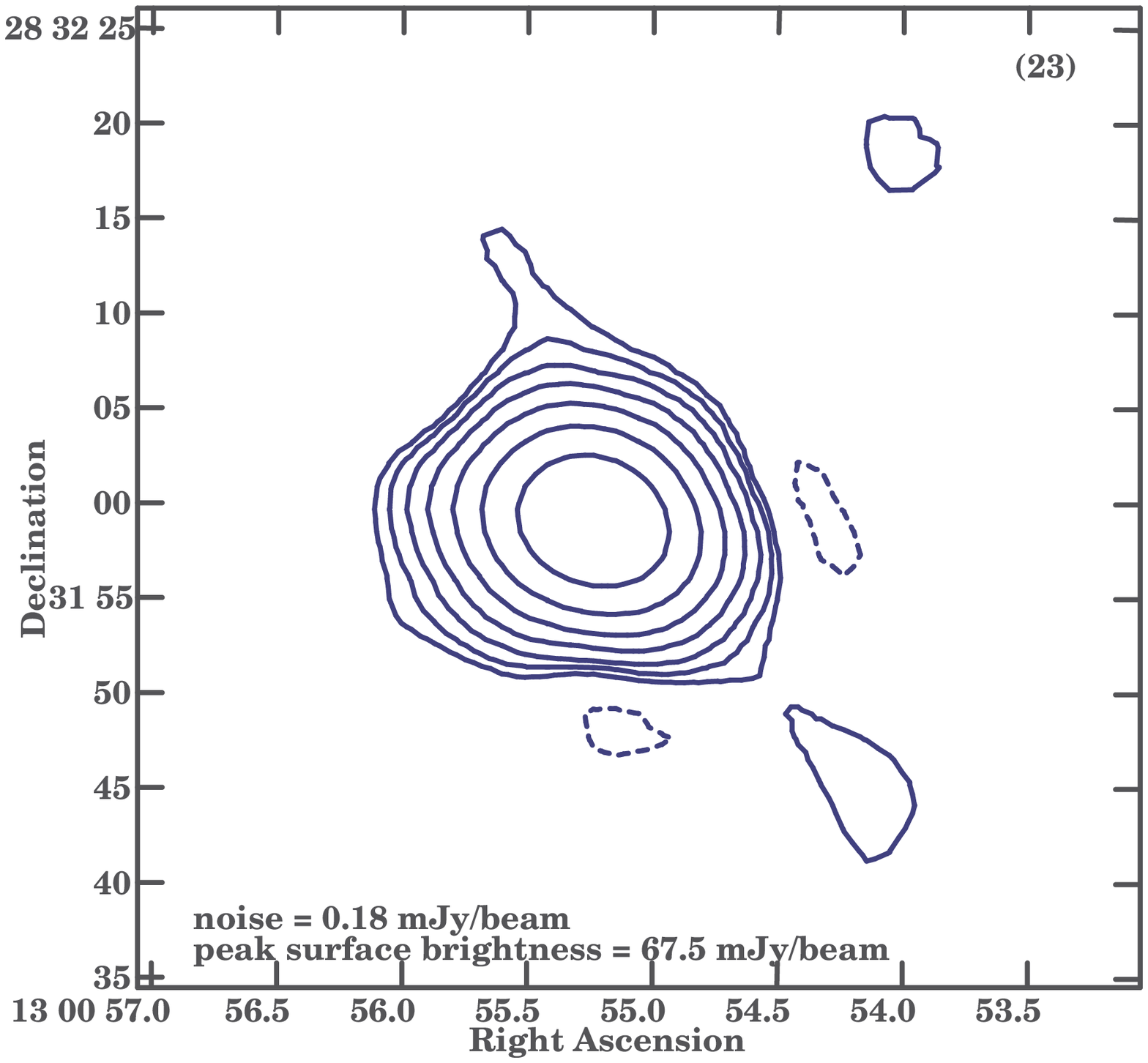} &
\includegraphics[height=3.93cm]{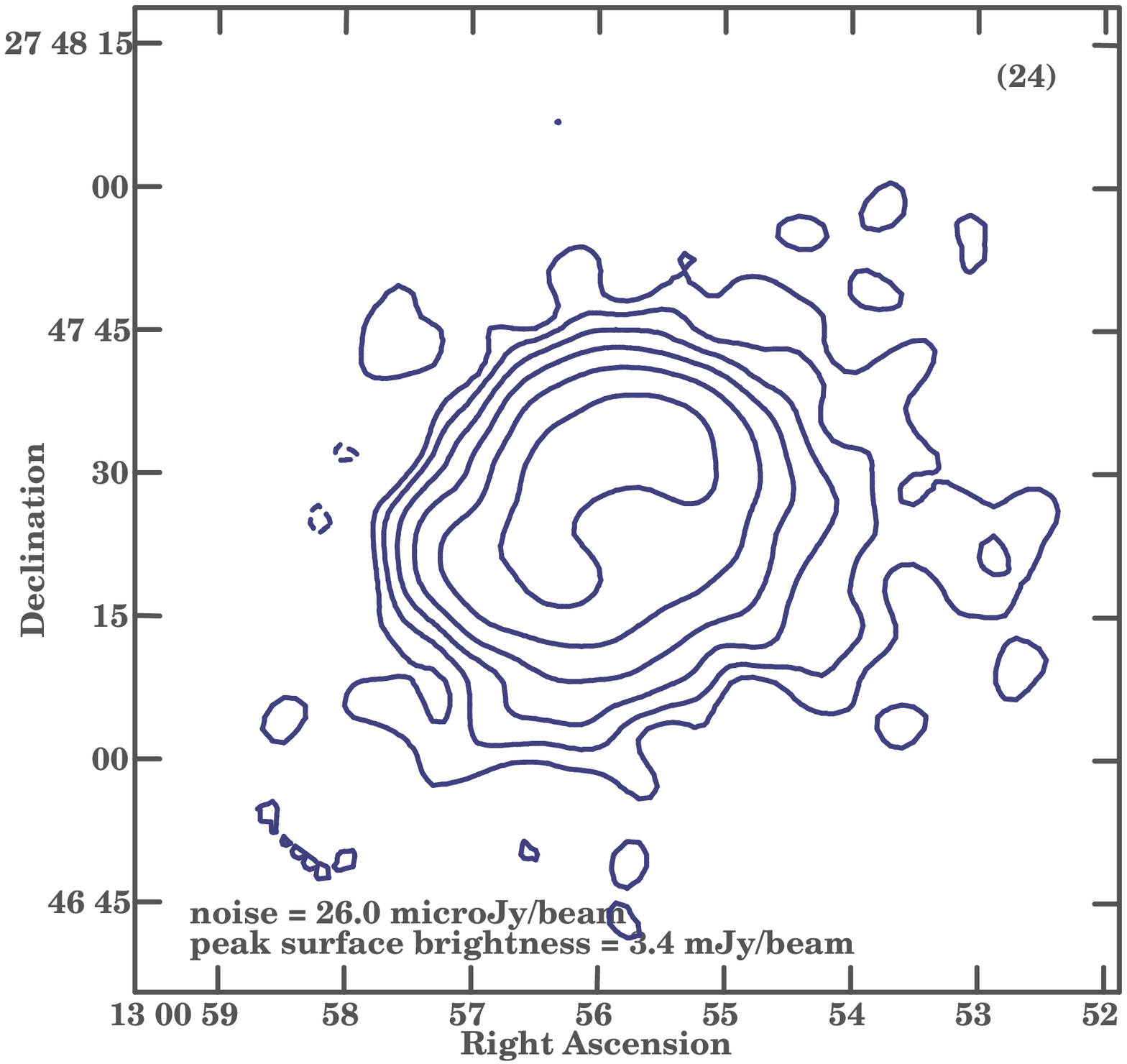} \\
\includegraphics[height=3.93cm]{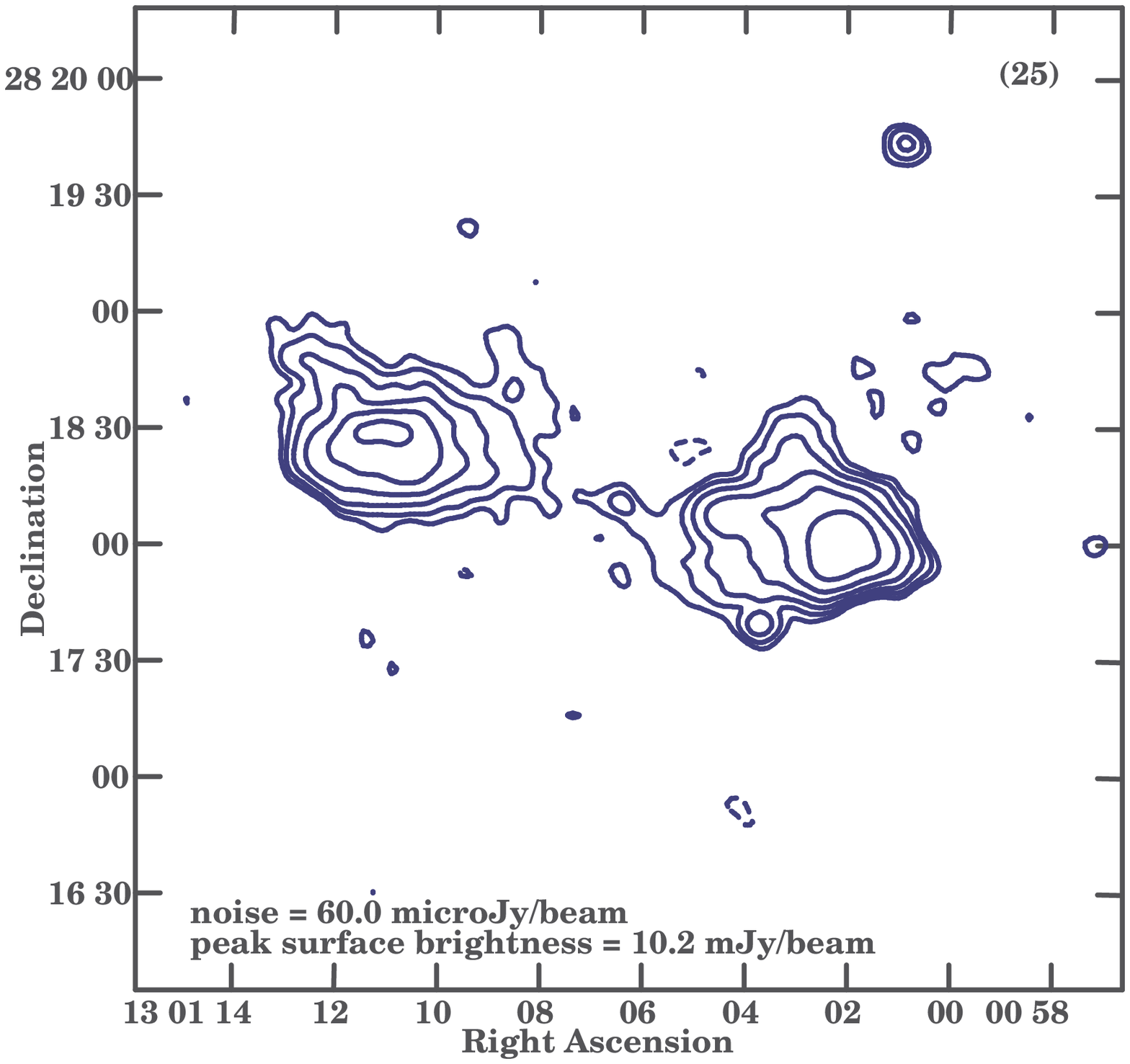} &
\includegraphics[height=3.93cm]{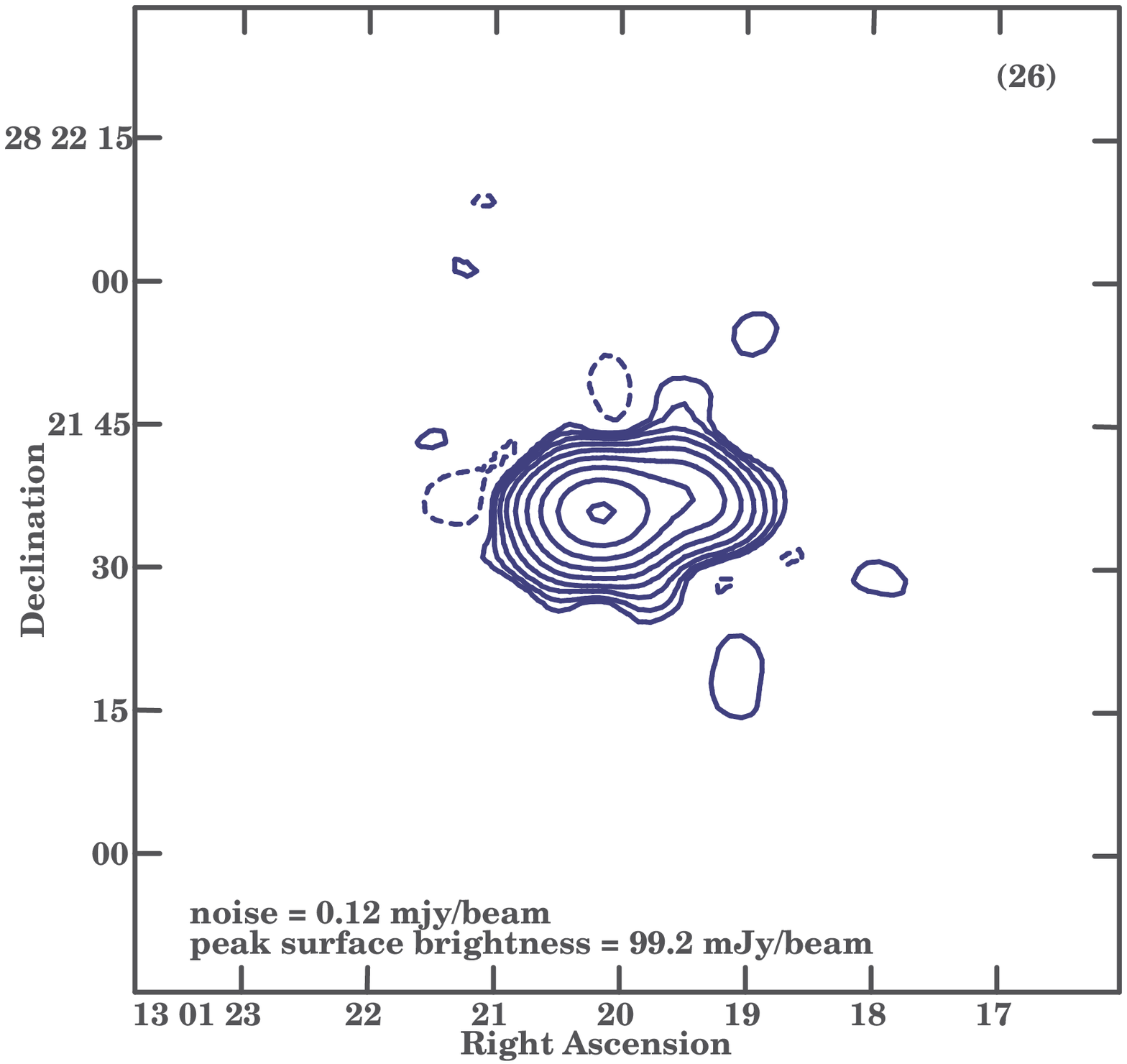} &
\includegraphics[height=3.93cm]{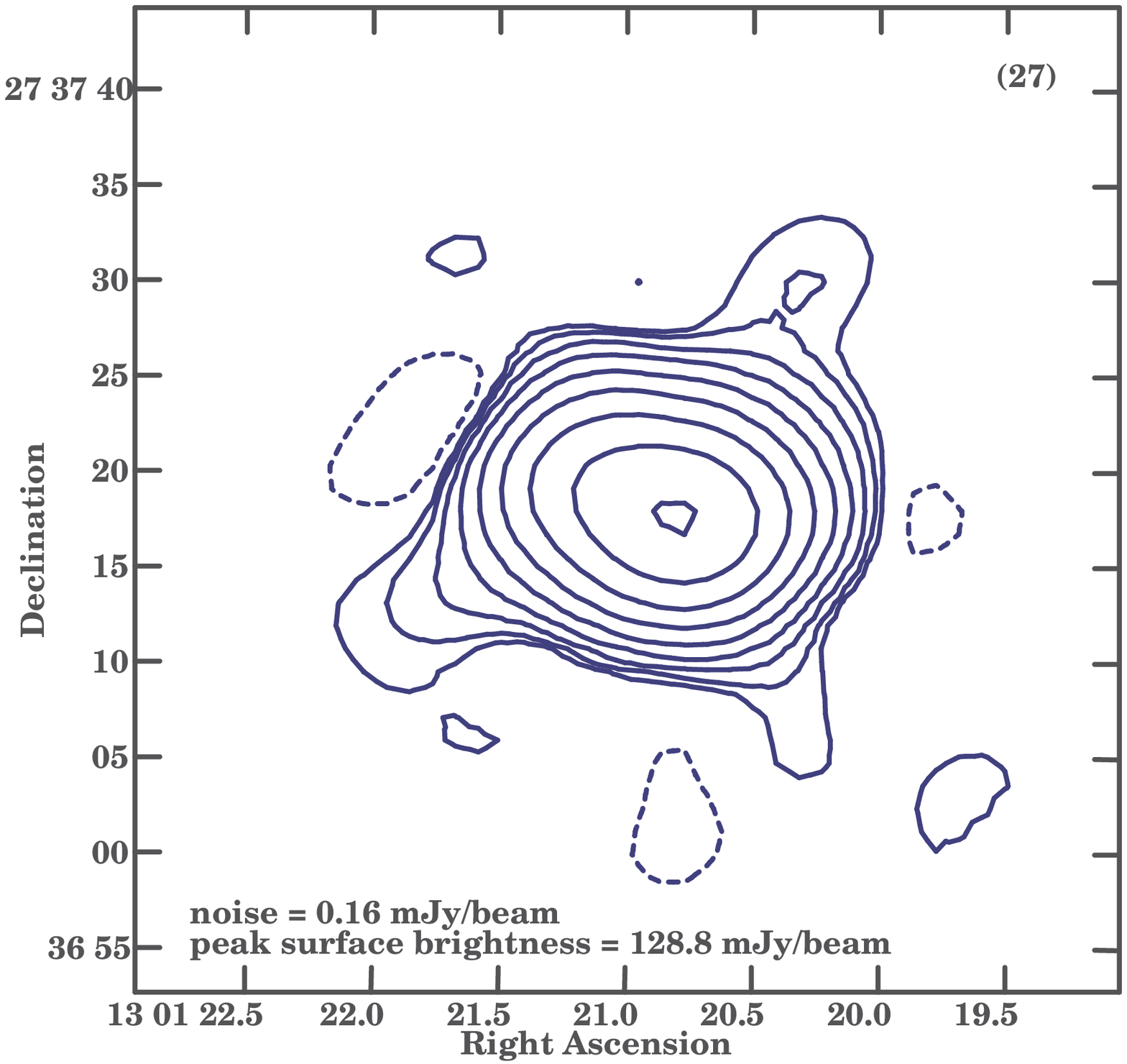} &
\includegraphics[height=3.93cm]{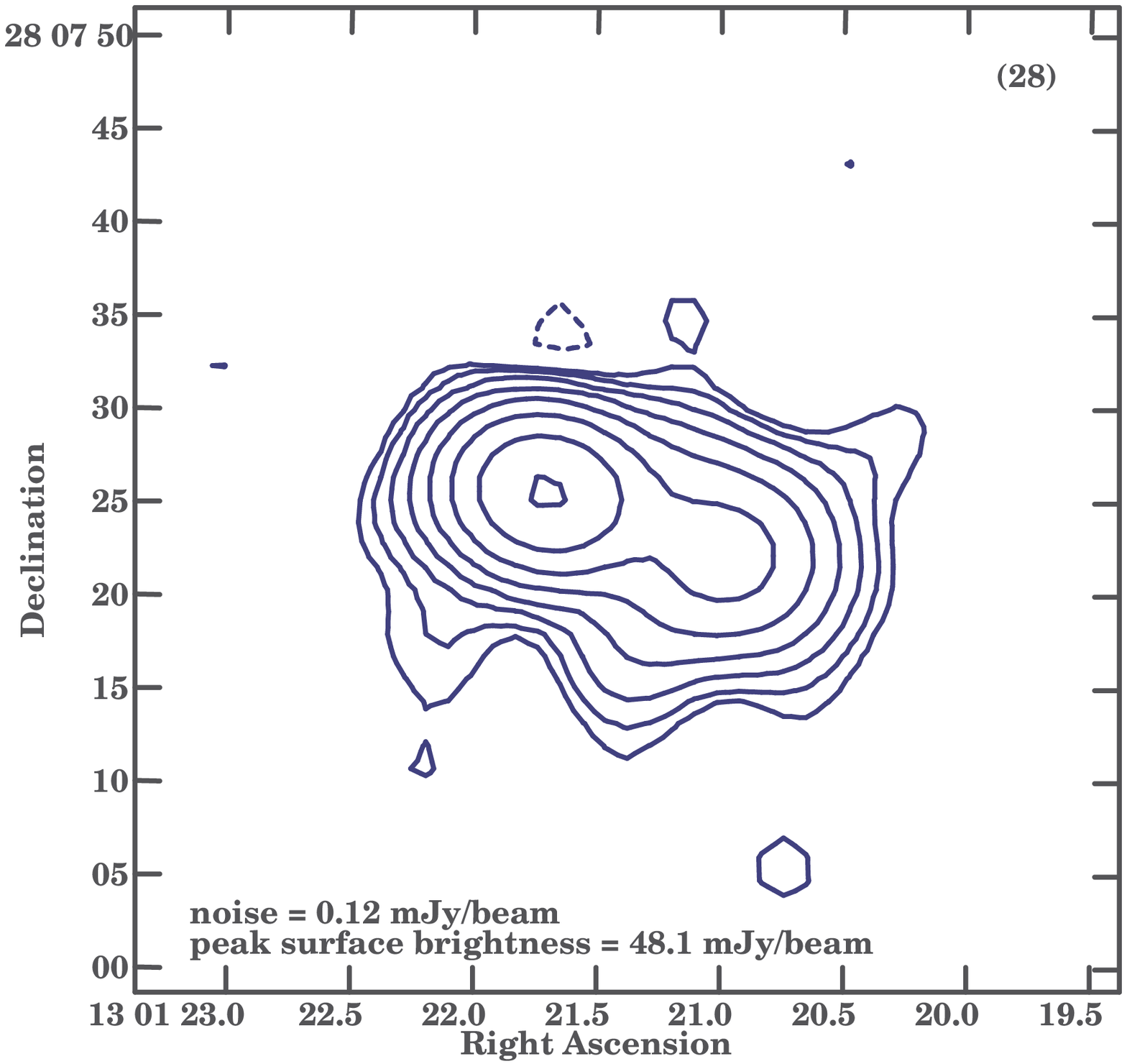} \\
\includegraphics[height=3.93cm]{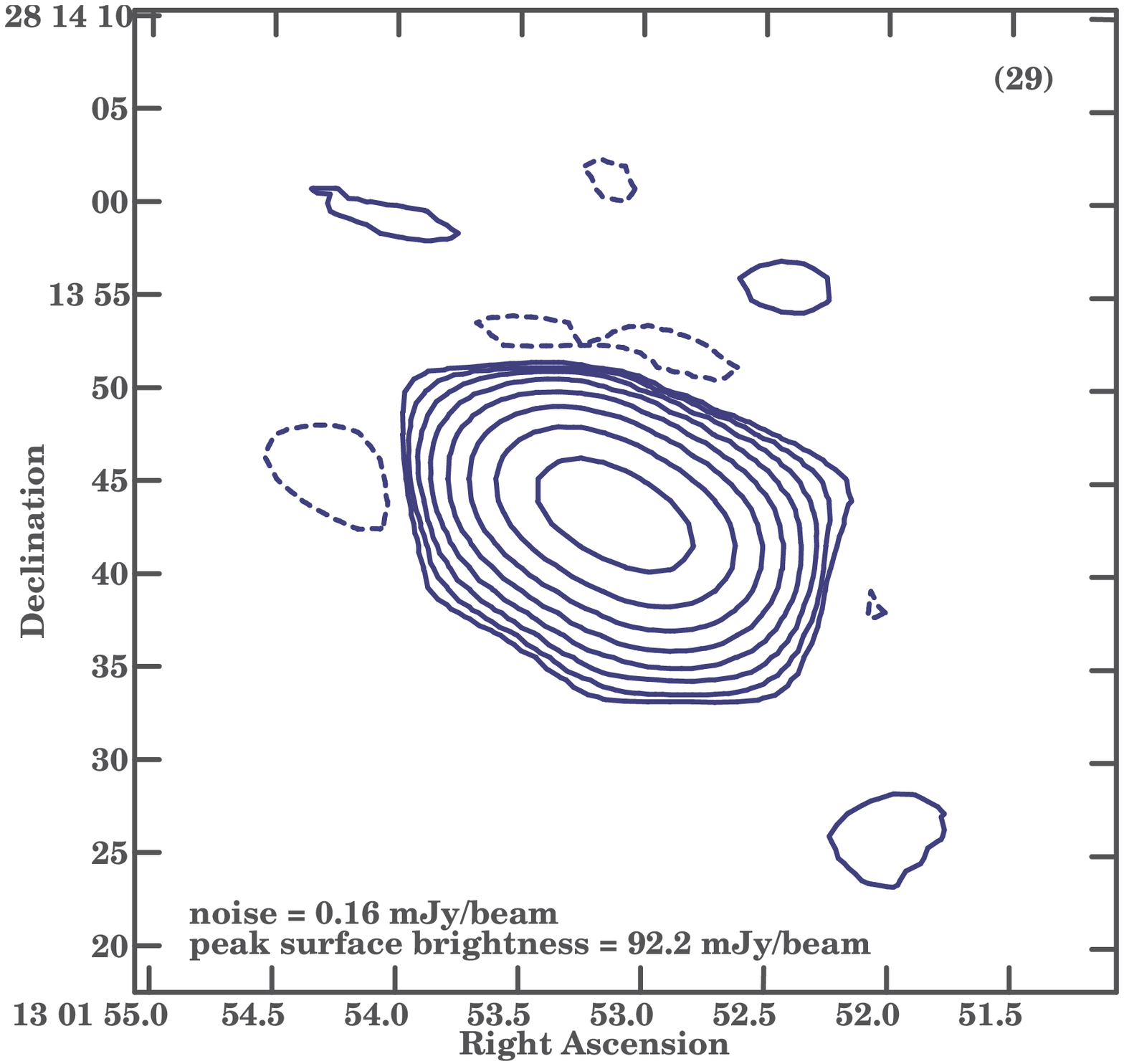} &
\includegraphics[height=3.93cm]{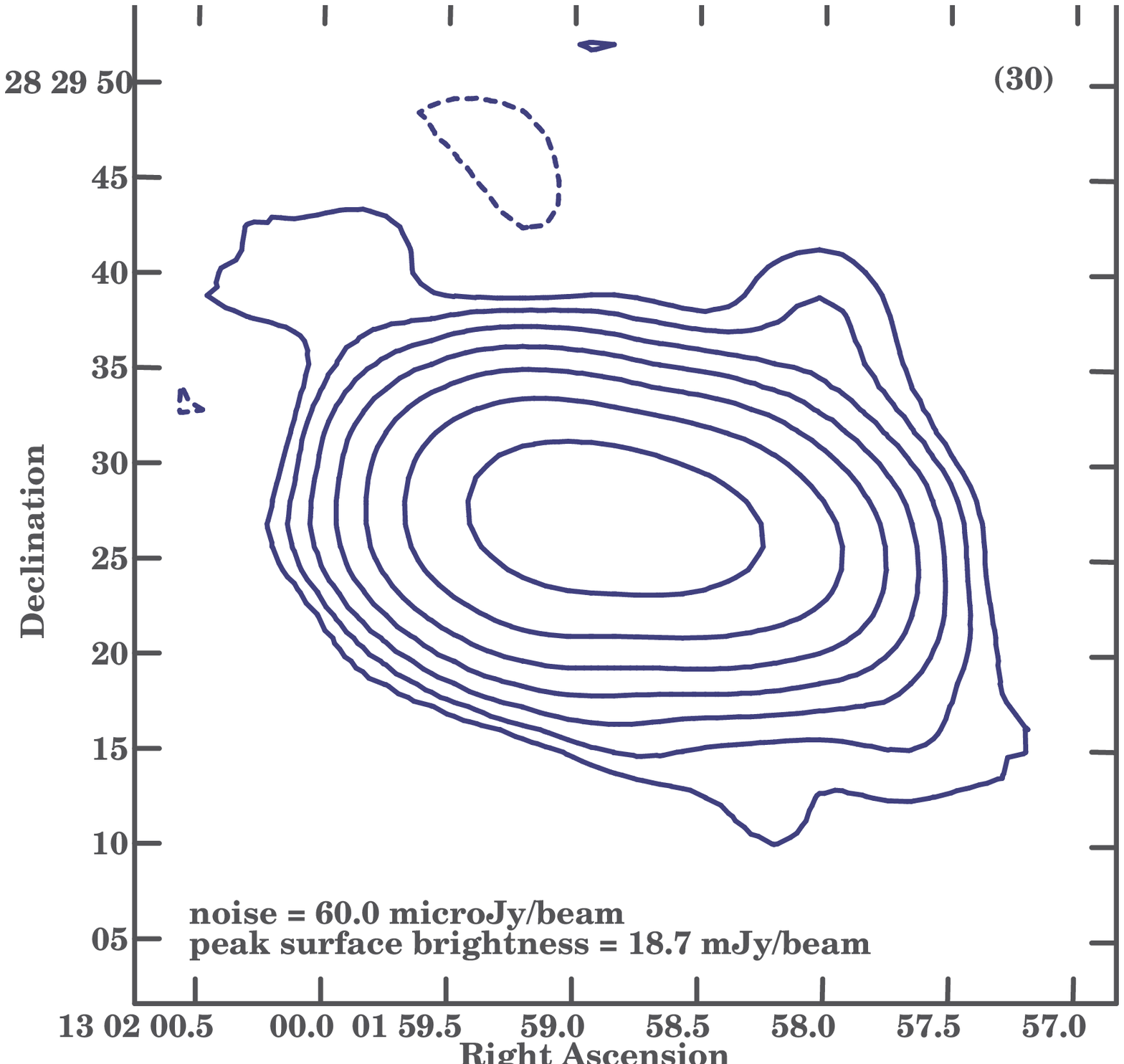} &
\includegraphics[height=3.93cm]{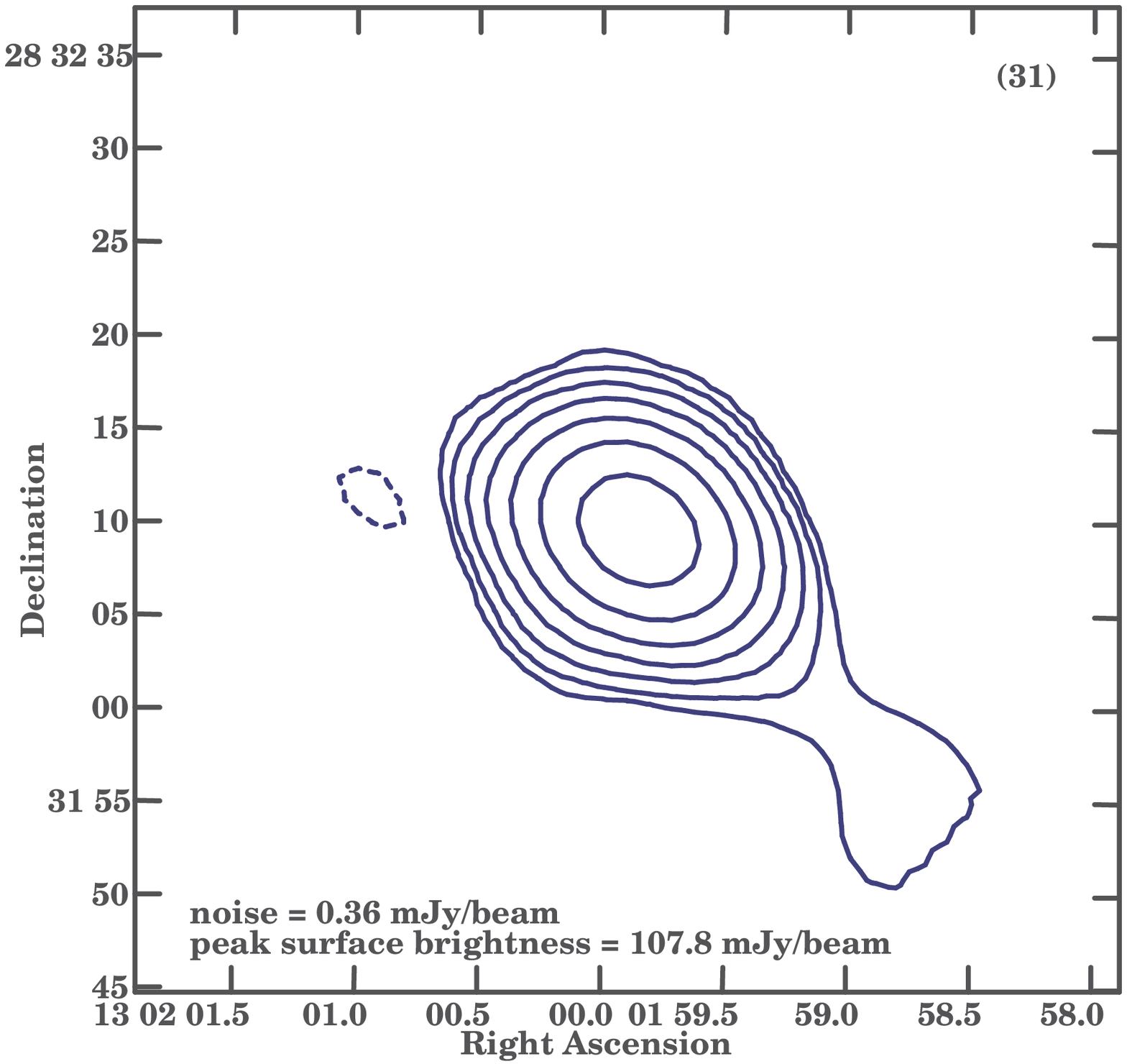} &
\includegraphics[height=3.93cm]{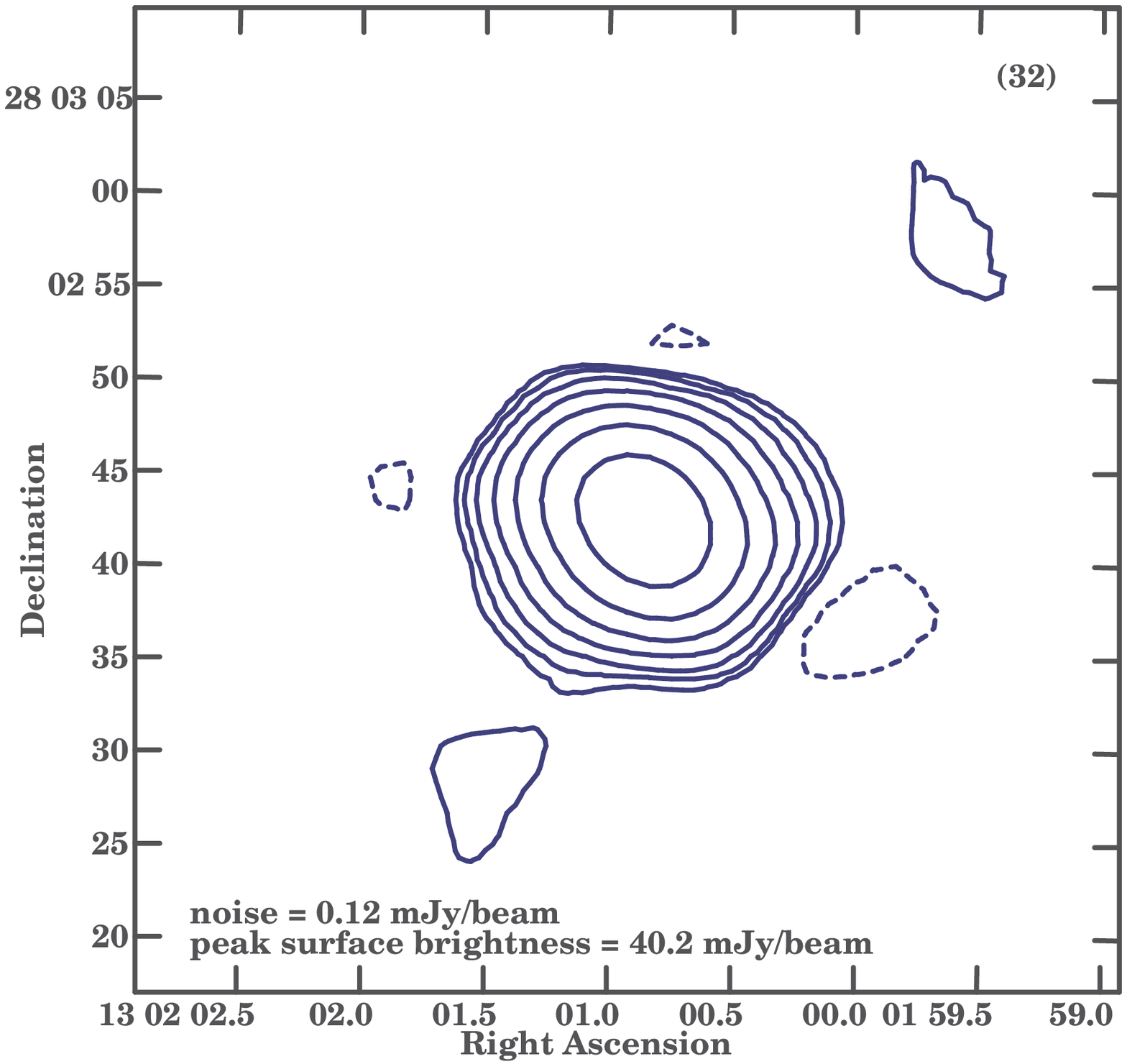}
\end{tabular}
\caption{Continued}
\end{center}
\end{figure*}

\item[02.] NGC\,4869 $-$
Another member of the Coma cluster lies well outside
the cluster core: $\sim$40$^{\prime}$ away to the southwest of NGC\,4874.
The radio source is hosted by an elliptical host \citep{1990AJ.....99.1381V}.
The structure of this source is typical of a head-tail, narrow angle tail \citep[NAT, e.g., 3C\,129:][]{2004A&A...420..491L} radio
source with a weak unresolved radio core, two oppositely directed radio jets,
and a long-low surface brightness tail pointing away from the cluster center.
The tailed-jet has a conical shape centered on the nucleus and it initially
expands and then recollimates.
Subsequently, the radio jet bends by $\sim$70~deg with respect to the initial direction of
propagation.  There is also the presence of diffuse extensions toward the inner
edge and sharpness in surface brightness toward the outer edge of the radio
source, possibly due to the motion of this head-tail radio source around
the dark matter potential.

Our 1050--1450\,MHz band uGMRT image again shows a head-tail radio source.  We detect the weak radio core, with two narrow angle radio jets pointing away from the radio core forming a long tail.  The conical shaped jet centered on the nucleus expands, recollimated and then bends by $\sim$70~deg as is seen in our uGMRT 250-500 MHz band image, which is also reported by \citet{Ferettietal1990}.  The high surface brightness outer edge is seen, whereas the low surface brightness, diffuse extensions toward the inner edge seen in our low-frequency 250--500 MHz band image are not seen in our high-frequency 1050--1450 MHz band image, probably because of synchrotron cooling \citep{Ferettietal1990}.

\item[03.] NGC\,4839 $-$
This is the compact dominant galaxy of the NGC\,4839 group and
lies in a peripheral Coma cluster region.  The optical host is a 
low/average surface brightness, disk-dominated galaxy \citep{Oemler1976}.
It is located at nearly the same $z$ $=$ 0.02456 of the Coma cluster.
A two-sided radio jet with slight distortions emerging in P.A. $\sim$0$^\circ$
with a bow shock-like a feature with indications of ram pressure stripping,
which suggests that the NGC\,4839 group is falling into the Coma cluster
\citep{2001A&A...365L..74N}.
The integrated flux density, $S_{\rm 408\,MHz}$ \citep[= 316.7 mJy;][]{Kimetal1994} is a factor of $\sim$1.6 higher than our measurement.  This is because the radio source is also a relic source lying at the edge of the field of view, where our ability to map low surface brightness diffuse emission is poor. 

\begin{figure*}[ht]
\begin{center}
\begin{tabular}{ccc}
\includegraphics[height=5.05cm]{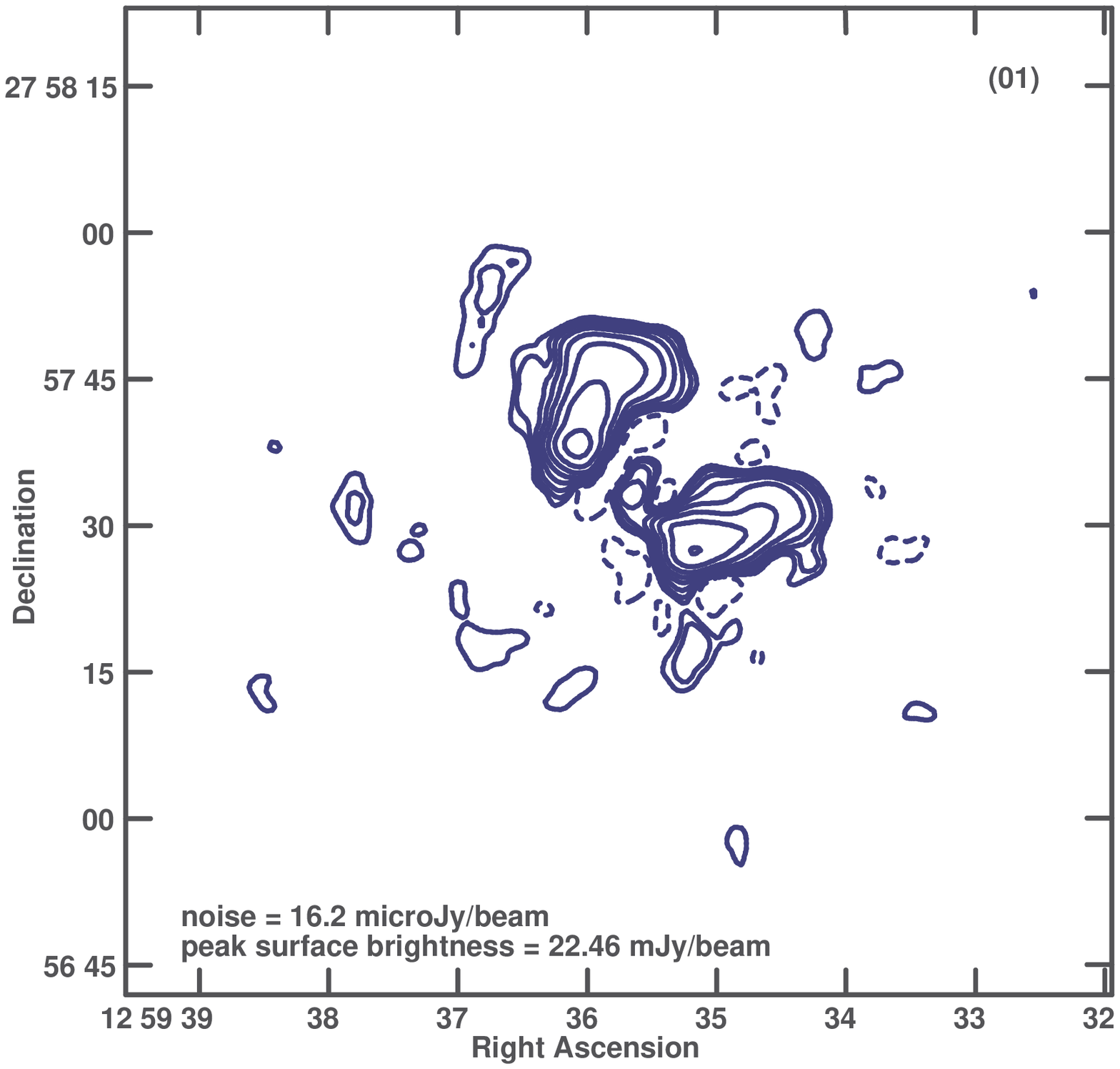} &
\includegraphics[height=5.05cm]{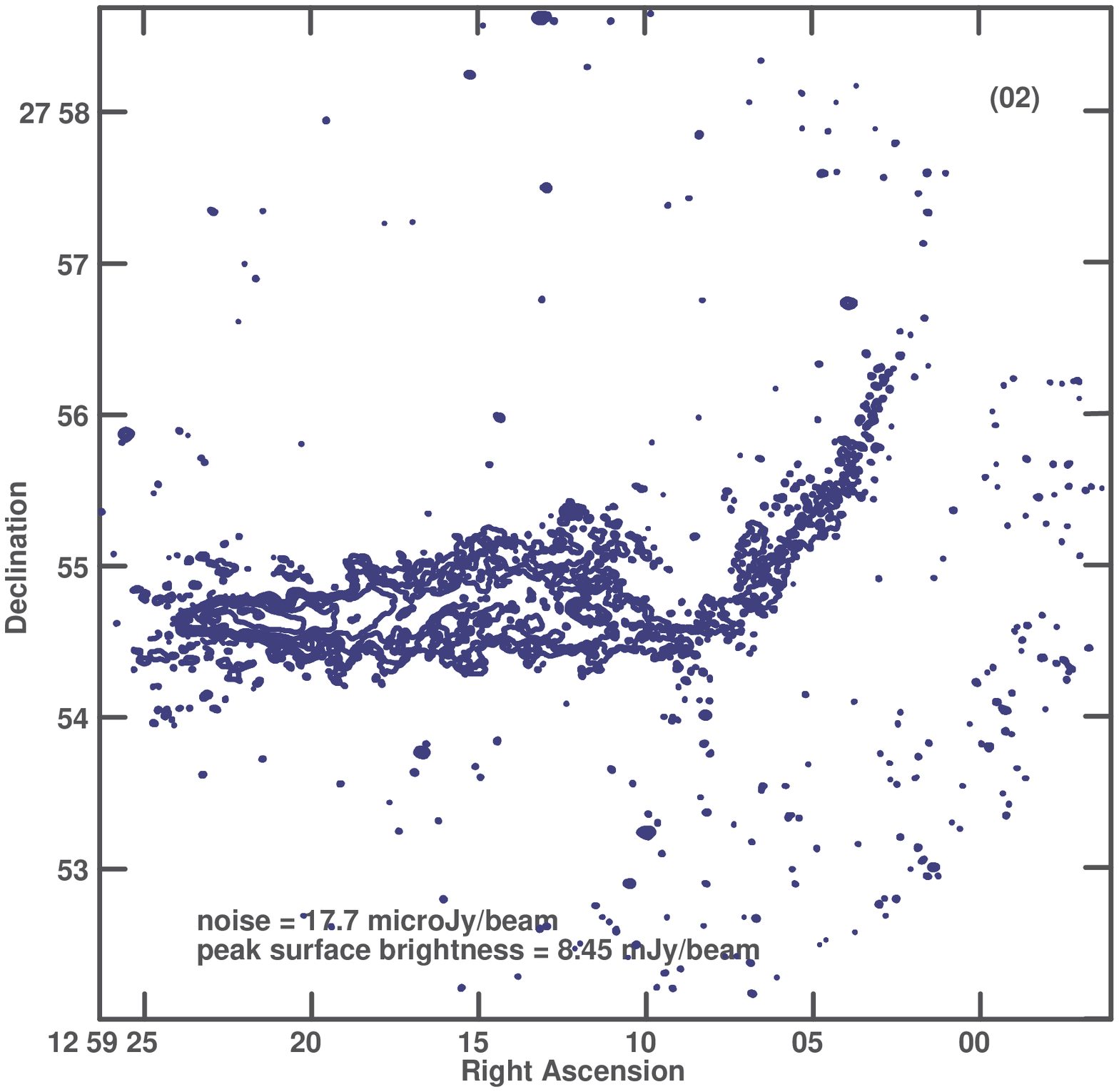} &
\includegraphics[height=5.05cm]{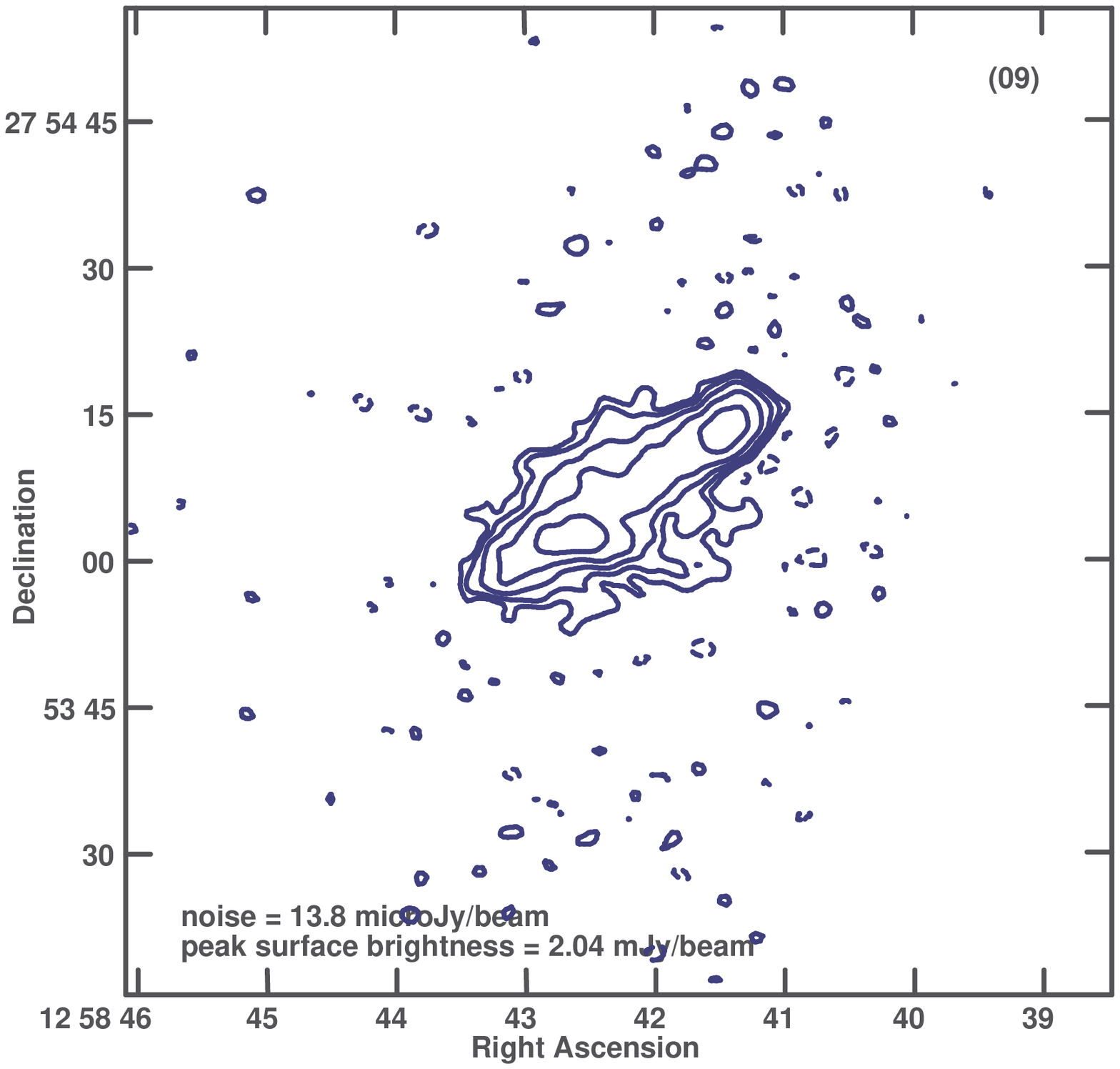} \\
\includegraphics[height=5.05cm]{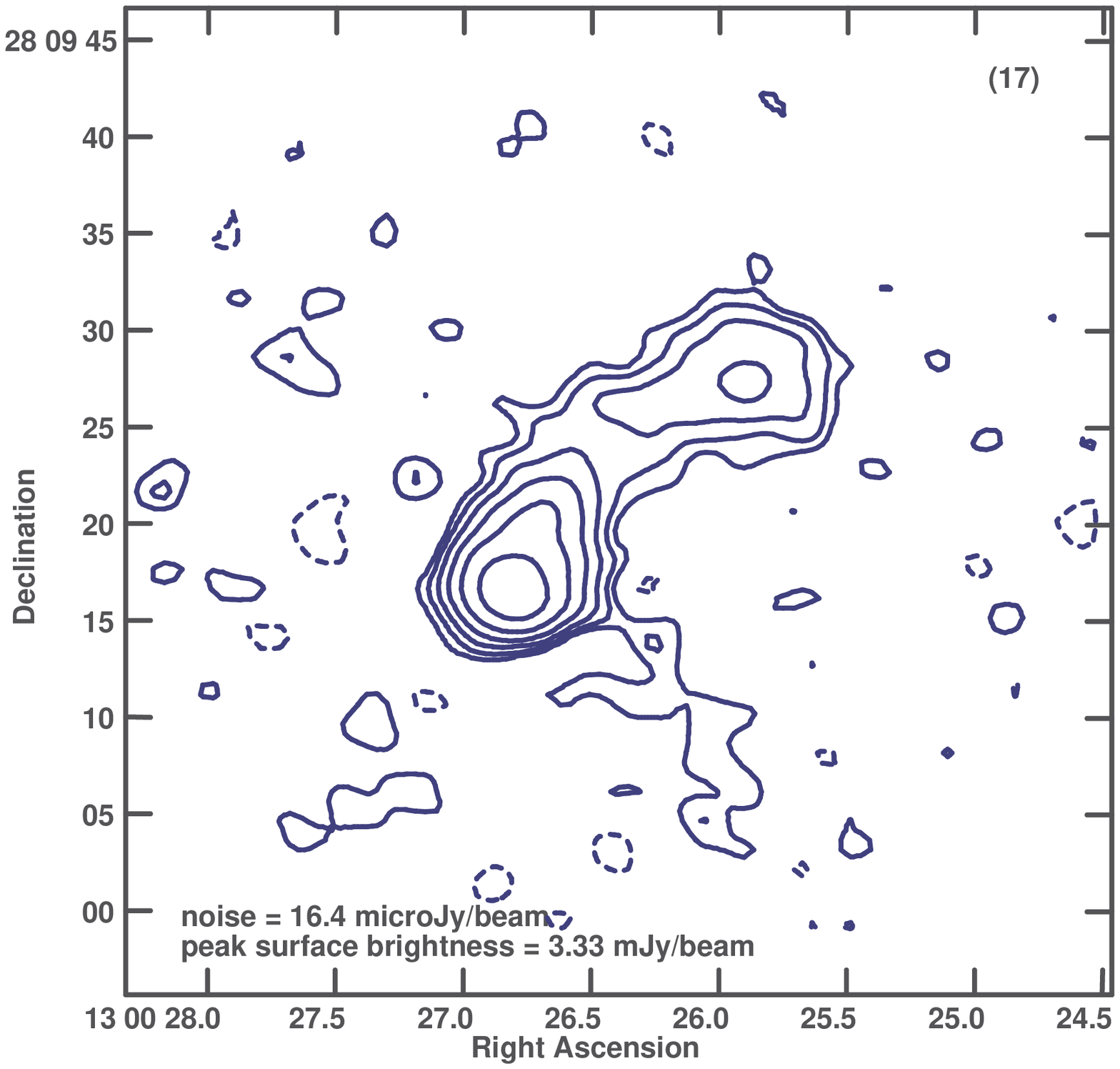} &
\includegraphics[height=5.05cm]{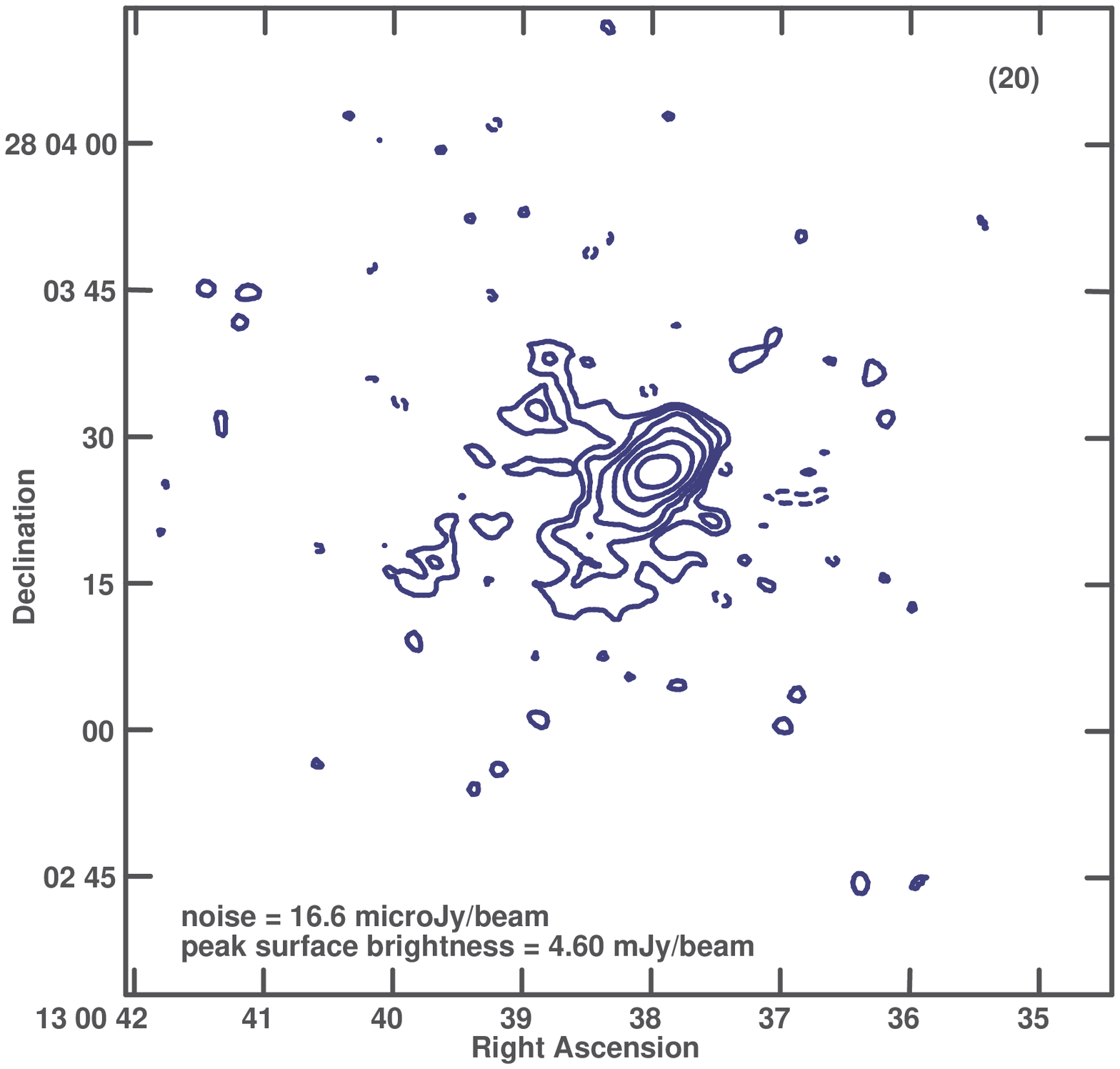} &
\end{tabular}
\caption{Images of the five of 32 brightest ($S_\nu \gtrsim$ 30~mJy) and dominant, both
pointlike and extended radio sources in the Coma cluster detected in a small field of view at the 1050--1450 MHz band of the uGMRT.
These radio sources are in the order presented in Table~\ref{tab:peel-model}.  The lowest radio contour plotted is three times the local \textsc{rms} noise and increasing by factors of 2.  The local \textsc{rms} noise and continuum peak surface brightness of the source are denoted in each panel (lower-left corner) along with its Source\_ID (upper-right corner).}
\label{collage-Lband}
\end{center}
\end{figure*}

\item[06.] NGC\,4848 $-$
A head-tail radio source appears to be undergoing ram pressure stripping \citep[see also][]{Chenetal}.  The optical host is a blue disk Scd galaxy
\citep{2000AJ....119..580B}.
\item[09.] 5C\,4.70 $-$
The uGMRT 1050--1450 MHz band image shows a classic double source FR\,II radio morphology \citep{FanaroffRiley} consisting of edge-brightened, diffuse lobe emission with clear hot spots at the end of the source.
\item[12.] WISEA\,J125859.32$+$274644.9 $-$
A classic two-sided double radio source of FR\,II radio morphology.  The radio core is well detected.  
\item[13.] SDSS\,J125903.80$+$281145.6 $-$
This again is a classic radio galaxy of FR\,II radio morphology,
though the southwest radio lobe is associated with 5C\,4.77 radio source.

\item[14.] WISEA\,J130006.87$+$272936.8 $-$
An unresolved radio source, which is $\sim$4$^{\prime\prime}$ away from Abell\,1656:[EF2011] 1595 (NED classification) source.  This NED classified source has a photometric redshift of 0.22001.

\tabletypesize{\scriptsize}
\def\arraystretch{0.5}
\begin{table*}
\tablewidth{0pt}
\centering
\caption{A short summary of the nature of dominant sources in the Coma cluster field at 400 MHz using the uGMRT.}
\begin{tabular}{clllrl}
\hline
 & Source ID & \multicolumn{2}{c}{Position at 400 MHz}& \multicolumn{1}{c}{$z$} & Salient feature of radio morphology \\
\cline {1-2}  & &\multicolumn{1}{c}{R.A.} & \multicolumn{1}{c}{Decl.}  &     & \\
  & (1) & \multicolumn{1}{c}{(2)} & \multicolumn{1}{c}{(3)} & \multicolumn{1}{c}{(4)} & (5) \\
\hline\noalign{\smallskip}
 01. & NGC\,4874                    & 12:59:35.71 & $+$27:57:33.37 & 0.02394 & The BCG of Coma cluster; 5C\,4.85, C1B$-$127 \\
 02. & NGC\,4869                    & 12:59:23.36 & $+$27:54:41.73 & 0.02288 & An NAT radio source; 5C\,4.81, C1B$-$903 \\
 03. & NGC\,4839                    & 12:57:24.36 & $+$27:29:52.14 & 0.02456 & The cD galaxy of NGC\,4839 group; 5C\,4.51, C1C$-$904 \\[6pt]
 04. & WISEA\,J125656.53$+$281101.4 & 12:56:57.84 & $+$28:11:07.32 &         & A discrete point source that is $\sim$19$^{\prime\prime}$ away from 5C\,4.45 \\
 05. & WISEA\,J125752.93$+$280609.4 & 12:57:52.74 & $+$28:06:08.26 & 0.45360 & An FR\,II radio galaxy; 5C\,4.54, C1B$-$012 \\
 06. & NGC\,4848                    & 12:58:05.54 & $+$28:14:32.30 & 0.02351 & An NAT radio source; 5C\,4.58, C1A$-$020 \& C1A$-$021  \\
 07. & SDSS\,J125811.97$+$282748.6  & 12:58:12.00 & $+$28:27:48.87 & 3.17688 & An unresolved radio source; C1A$-$026 \\
 08. & WISEA\,J125812.36$+$273534.7 & 12:58:12.26 & $+$27:35:34.11 & 0.44673 & An FR\,II radio galaxy; C1C$-$042, C1C$-$047 \& C1C$-$049 \\
 09. & 5C\,4.70                     & 12:58:42.24 & $+$27:54:06.17 &         & A double radio source; C1B$-$065 \\
 10. & WISEA\,J125854.14$+$281752.3 & 12:58:54.12 & $+$28:17:52.78 & 0.38160 & Point-like, unresolved radio source; C1A$-$056 \\
 11. & WISEA\,J125858.10$+$281924.6 & 12:58:58.08 & $+$28:19:27.41 &         & A discrete point source; 5C\,4.75, C1A$-$060 \\
 12. & WISEA\,J125859.32$+$274644.9 & 12:58:59.52 & $+$27:46:45.00 & 0.02300 & An FR\,II radio galaxy; 5C\,4.74, C1C$-$61 \\
 13. & SDSS\,J125903.80$+$281145.6  & 12:59:03.43 & $+$28:12:01.46 &         & An FR\,II radio galaxy; C1A$-$061 \\
 14. & WISEA\,J130006.87$+$272936.8 & 13:00:06.84 & $+$27:29:37.33 &         & An unresolved source; C1C$-$151 \\
 15. & SDSS\,J130014.76$+$273932.8  & 13:00:14.76 & $+$27:39:34.39 & 0.37984 & Amorphous, high surface brightness radio source \\
 16. & NVSS\,J130023$+$271915       & 13:00:23.76 & $+$27:19:14.51 &         & A discrete point and an unresolved source; 5C\,4.97 \\
 17. & WISEA\,J130026.68$+$280920.0 & 13:00:26.35 & $+$28:09:21.74 &         & A double radio source; 5C\,4.102, C1B$-$187 \\
 18. & FBQS J1300$+$2830            & 13:00:28.44 & $+$28:30:10.34 & 0.64709 & A pointlike radio quasar source; 5C\,4.105, C1A$-$138 \\
 19. & SDSS\,J130030.76$+$272239.4  & 13:00:31.12 & $+$27:22:36.31 &         & A bent wide-angle  (triple) radio source \\
 20. & IC\,4040                     & 13:00:38.16 & $+$28:03:24.25 & 0.02615 & An NAT radio source; 5C\,4.108, C1B$-$904 \\
 21. & WISEA\,J130043.86$+$282458.8 & 13:00:42.98 & $+$28:24:55.23 & 0.02095 & A WAT radio source; 5C\,4.113, C1A$-$152, C1A$-$153, C1A$-$156, \\
     &                              &             &                &         & C1A$-$922 \& C1A$-$923 \\
 22. & WISEA\,J130050.84$+$280803.6 & 13:00:50.88 & $+$28:08:04.07 &         & A double radio source with core-jet morphology; 5C\,4.114 \\
 23. & SDSS\,J130055.21$+$283205.7  & 13:00:55.08 & $+$28:32:00.41 &         & A discrete point source \\
 24. & NGC\,4911                    & 13:00:56.08 & $+$27:47:27.02 & 0.02663 & An amorphous, diffuse radio source; 5C\,4.117, C1B$-$901 \\
 25. & WISEA\,J130106.38$+$281813.0 & 13:01:06.27 & $+$28:18:11.57 & 0.37339 & A radio galaxy; 5C\,4.122, C1A$-$169, C1A$-$910 \& C1A$-$911 \\
 26. & $[$HB89$]$ 1258$+$286 NED02  & 13:01:19.92 & $+$28:21:37.58 & 1.37176 & A pointlike radio source; 5C\,4.127, C1A$-$177 \& C1A$-$178 \\
 27. & WISEA\,J130120.88$+$273718.4 & 13:01:21.00 & $+$27:37:18.56 &         & An unresolved source; 5C\,4.128, C1C$-$205 \& C1C$-$206 \\
 28. & WISEA\,J130121.84$+$280726.7 & 13:01:21.36 & $+$28:07:25.62 & 2.26500 & A double radio source; 5C\,4.129, C1B$-$244 \& C1B$-$245 \\
 29. & WISEA\,J130153.05$+$281344.0 & 13:01:53.04 & $+$28:13:44.49 &         & An unresolved source; 5C\,4.134, C1A$-$187 \& C1A$-$188 \\
 30. & WISEA\,J130158.70$+$282928.7 & 13:01:58.80 & $+$28:29:28.56 &         & A pointlike dumbbell radio source; 5C\,4.137 \\
 31. & WISEA\,J130159.77$+$283210.1 & 13:01:59.88 & $+$28:32:11.14 &         & A pointlike source with an extension; 5C\,4.138 \\
 32. & WISEA\,J130200.78$+$280247.3 & 13:02:00.96 & $+$28:02:43.45 &         & A discrete pointx and an, unresolved source; 5C\,4.140, C1B$-$276 \\
\hline
\end{tabular}
\label{tab:peel-model}
\tablecomments{Notes on our interpretation of the radio morphologies for these brightest and dominant sources are discussed in Sec.~\ref{notes-dominant}. \\
Col.~1: Source name as identified in the NED. \\
Col.~2 and 3: The R.A. and decl. (J2000); barring first three sources, the rest of the sources are ordered by the increasing R.A.  Source position is given by the brightest component for the unresolved source, whereas for the radio double or the diffuse source, the position is given by the possible radio core. \\
Col.~4: spectroscopic redshift of source. \\
Col.~5: a comment on the morphology of the radio source at the 250--500~MHz band.  Wherever possible, we also list their 5C \citep[fifth Cambridge;][]{Willson1970} survey catalog names and unique `C1A/B/C' source IDs that are reported in \citet{MHM2009}.}
\end{table*}

\item[17.] WISEA\,J130026.68$+$280920.0 $-$
The uGMRT 1050--1450 MHz band image shows a double source of FR\,II radio morphology, which consists of an edge-brightened, diffuse radio lobe emission with clear hot spots at the end of the source. The two jets have bent after emanating from the radio core, probably because of ram pressure.
\item[19.] SDSS\,J130030.76$+$272239.4 $-$
We have classified it to be a triple source because the 2.5 times the \textsc{rms} noise contour shows three bright peaks as a single connected contiguous radio source
and the SDSS optical source is located at the center.  If our understanding
is correct, it is possibly a WAT radio galaxy.
\item[20.] IC\,4040 $-$
A NAT radio source hosted by CGCG\,160$-$252 galaxy of spiral or an irregular morphological type.
It appears to be undergoing ram pressure stripping \citep[see also][]{Chenetal}.

This NAT source does not show a similar extent in the 1050--1450 MHz band image as compared to the 250--500 MHz band image, probably because of synchrotron cooling.

\tabletypesize{\scriptsize}
\def\arraystretch{0.5}
\begin{table*}
\tablewidth{0pt}
\centering
\caption{{\scriptsize {The total intensity and spectral index for all dominant sources the Coma cluster field along with radio luminosity for 17 sources with known spectroscopic redshifts.}}}
\begin{tabular}{clrrrrcccr}
\hline
 & Source ID & $S_{\rm 151\,MHz}$ & $S_{\rm 400\,MHz}$ & $S_{\rm 408\,MHz}$ & $S_{\rm 1250\,MHz}$ & $S_{\rm 1400\,MHz}$ & $S_{\rm 4850\,MHz}$ & $\alpha$ & L$_{\rm 400\,MHz}$ \\
  & & \multicolumn{1}{c}{(mJy)} & \multicolumn{1}{c}{(mJy)} & \multicolumn{1}{c}{(mJy)} & \multicolumn{1}{c}{(mJy)} & \multicolumn{1}{c}{(mJy)} & \multicolumn{1}{c}{(mJy)} & \multicolumn{2}{r}{(erg~s$^{-1}$~Hz$^{-1}$)} \\
  \multicolumn{2}{c}{(1)}& \multicolumn{1}{c}{(2)} & \multicolumn{1}{c}{(3)} & \multicolumn{1}{c}{(4)} & \multicolumn{1}{c}{(5)} & \multicolumn{1}{c}{(6)} & \multicolumn{1}{c}{(7)} & \multicolumn{1}{c}{(8)} & \multicolumn{1}{c}{(9)} \\
\hline\noalign{\smallskip}
 01. & NGC\,4874           &  900 $\pm$81$^{\rm b}$   &  476.6 $\pm$2.3 & 454.4 $\pm$22.7 & 296.9 $\pm$2.3 & 206.3 $\pm$0.2$^{\rm a}$ &  84.0 $\pm$12.6$^{\rm d}$ & $-$0.67 $\pm$0.08 & 6.2 $\times$ 10$^{30}$ \\
 02. & NGC\,4869           & 2780 $\pm$208$^{\rm b}$  & 1381.0 $\pm$5.7 &1387.0 $\pm$69.4 & 584.1 $\pm$3.6 & 402.0 $\pm$0.7$^{\rm a}$ & 108.0 $\pm$16.2$^{\rm d}$ & $-$0.99 $\pm$0.20 & 1.7 $\times$ 10$^{31}$ \\
 03. & NGC\,4839           &  660 $\pm$71.9$^{\rm b}$ &  196.9 $\pm$2.1 & 316.7 $\pm$10.4 & \nodata        &  77.0 $\pm$0.5$^{\rm a}$ &  33.0 $\pm$~5.0$^{\rm d}$ & $-$0.73 $\pm$0.07 & 3.0 $\times$ 10$^{30}$ \\[6pt]
 04. & WISEA\,J1256$+$2811 &  \nodata                 &   75.9 $\pm$0.2 &  \nodata        & \nodata        &  10.4 $\pm$0.5$^{\rm e}$ &  \nodata                  & $-$1.59 $\pm$0.02 & \\
 05. & WISEA\,J1257$+$2806 &  390 $\pm$58.5$^{\rm b}$ &  206.7 $\pm$0.7 & 205.0 $\pm$7.3~ & \nodata        &  75.3 $\pm$0.4$^{\rm a}$ &  30.0 $\pm$~4.5$^{\rm d}$ & $-$0.79 $\pm$0.05 & 1.5 $\times$ 10$^{33}$ \\
 06. & NGC\,4848           &   90 $\pm$12.0$^{\rm c}$ &   57.4 $\pm$0.3 &  54.2 $\pm$5.0~ & \nodata        &  14.7 $\pm$0.2$^{\rm a}$ &  \nodata                  & $-$1.09 $\pm$0.02 & 0.7 $\times$ 10$^{30}$ \\
 07. & SDSS\,J1258$+$2827  &  \nodata                 &  311.8 $\pm$0.2 &  \nodata        & \nodata        &  88.9 $\pm$0.4$^{\rm a}$ &  \nodata                  & $-$1.00 $\pm$0.02 & 2.9 $\times$ 10$^{35}$ \\
 08. & WISEA\,J1258$+$2735 &  \nodata                 &   37.8 $\pm$0.3 &  \nodata        & \nodata        &  15.8 $\pm$0.3$^{\rm a}$ &  \nodata                  & $-$0.70 $\pm$0.02 & 2.5 $\times$ 10$^{32}$ \\
 09. & 5C\,4.70            &  310 $\pm$69.4$^{\rm b}$ &  134.8 $\pm$0.4 &  \nodata        &  53.1 $\pm$0.2 &  32.9 $\pm$0.2$^{\rm a}$ &  \nodata                  & $-$1.13 $\pm$0.06 & \\
 10. & WISEA\,J1258$+$2817 &  490 $\pm$88.7$^{\rm b}$ &  210.0 $\pm$0.5 &  \nodata        & \nodata        &  74.8 $\pm$0.2$^{\rm a}$ &  35.0 $\pm$~5.3$^{\rm d}$ & $-$0.79 $\pm$0.08 & 1.1 $\times$ 10$^{33}$ \\
 11. & WISEA\,J1258$+$2819 &  \nodata                 &   82.6 $\pm$0.2 &  68.0 $\pm$6.8~ & \nodata        &  20.4 $\pm$0.1$^{\rm a}$ &  \nodata                  & $-$1.12 $\pm$0.02 & \\
 12. & WISEA\,J1258$+$2746 &  590 $\pm$88.5$^{\rm b}$ &  233.9 $\pm$0.7 &  \nodata        & \nodata        &  69.4 $\pm$0.2$^{\rm a}$ &   15.6 $\pm$0.8$^{\rm h}$     & $-$1.00 $\pm$0.08 & 2.8 $\times$ 10$^{30}$ \\
 13. & SDSS\,J1259$+$2811  &  \nodata                 &   54.7 $\pm$0.2 &  \nodata        & \nodata        &  12.9 $\pm$0.1$^{\rm a}$ &  \nodata                  & $-$1.15 $\pm$0.02 & \\
 14. & WISEA\,J1300$+$2729 &  \nodata                 &  138.1 $\pm$0.7 &  \nodata        & \nodata        &  58.4 $\pm$0.2$^{\rm a}$ &  \nodata                  & $-$0.69 $\pm$0.02 & \\
 15. & SDSS\,J1300$+$2739  &  \nodata                 &   49.1 $\pm$0.2 &  \nodata        & \nodata        &  21.9 $\pm$0.5$^{\rm e}$ &  \nodata                  & $-$0.64 $\pm$0.02 & 2.2 $\times$ 10$^{32}$ \\
 16. & NVSS\,J1300$+$2719  &  \nodata                 &  103.9 $\pm$0.2 & 124.6 $\pm$28.1 & \nodata        &  28.6 $\pm$1.3$^{\rm e}$ &  \nodata                  & $-$1.03 $\pm$0.02 & \\
 17. & WISEA\,J1300$+$2809 &   50 $\pm$12.0$^{\rm c}$ &   29.9 $\pm$0.3 &  \nodata        &  20.4 $\pm$0.1 &  15.3 $\pm$0.1$^{\rm a}$ &  \nodata                  & $-$0.53 $\pm$0.02 & \\
 18. & FBQS J1300$+$2830   &   70 $\pm$12.0$^{\rm c}$ &   91.2 $\pm$0.2 & 103.6 $\pm$12.7 & \nodata        & 117.2 $\pm$0.4$^{\rm a}$ & 188.0 $\pm$28.2$^{\rm d}$ & $+$0.31 $\pm$0.04 & 8.7 $\times$ 10$^{32}$ \\
 19. & SDSS\,J1300$+$2722  &  \nodata                 &   76.6 $\pm$0.3 &  \nodata        & \nodata        &  37.9 $\pm$0.5$^{\rm e}$ &  \nodata                  & $-$0.56 $\pm$0.02 & \\
 20. & IC\,4040            &  110 $\pm$12.0$^{\rm c}$ &   53.3 $\pm$0.1 &  48.3 $\pm$5.0~ &  24.7 $\pm$0.2 &  18.3 $\pm$1.6$^{\rm a}$ &  \nodata                  & $-$0.85 $\pm$0.02 & 0.8 $\times$ 10$^{30}$ \\
 21. & WISEA\,J1300$+$2824 &   60 $\pm$12.0$^{\rm c}$ &   31.3 $\pm$0.1 &  \nodata        & \nodata        &   9.4 $\pm$0.2$^{\rm a}$ &  \nodata                  & $-$0.96 $\pm$0.02 & 0.3 $\times$ 10$^{30}$$^{\rm f}$ \\
 22. & WISEA\,J1300$+$2808 &  240 $\pm$12.0$^{\rm c}$ &  118.1 $\pm$0.3 & 113.7 $\pm$11.4 & \nodata        &  48.6 $\pm$1.9$^{\rm e}$ &  14.9 $\pm$0.7$^{\rm h}$       & $-$0.77 $\pm$0.04 & \\
 23. & SDSS\,J1300$+$2832  &  \nodata                 &  100.1 $\pm$0.2 &  \nodata        & \nodata        &  40.8 $\pm$0.5$^{\rm e}$ &  \nodata                  & $-$0.72 $\pm$0.02 & \\
 24. & NGC\,4911           &   70 $\pm$12.0$^{\rm c}$ &   50.6 $\pm$0.2 &  55.7 $\pm$15.3 & \nodata        &  19.1 $\pm$0.3$^{\rm a}$ &  \nodata                  & $-$0.39 $\pm$0.02 & 0.5 $\times$ 10$^{30}$ \\
 25. & WISEA\,J1301$+$2818 &  350 $\pm$71.1$^{\rm b}$ &  209.7 $\pm$0.4 & 197.7 $\pm$10.5 & \nodata        &  79.5 $\pm$0.6$^{\rm a}$ &  \nodata                  & $-$0.77 $\pm$0.08 & 1.0 $\times$ 10$^{34}$ \\
 26. &$[$HB89$]$ 1258$+$286&  350 $\pm$54.6$^{\rm b}$ &  182.1 $\pm$0.6 & 173.9 $\pm$11.8 & \nodata        &  84.7 $\pm$0.1$^{\rm a}$ &  73.0 $\pm$11.0$^{\rm d}$ & $-$0.44 $\pm$0.07 & 1.3 $\times$ 10$^{34}$ \\
 27. & WISEA\,J1301$+$2737 &  \nodata                 &  207.6 $\pm$0.5 & 204.7 $\pm$20.5 & \nodata        &  70.5 $\pm$0.1$^{\rm a}$ &  \nodata                  & $-$0.86 $\pm$0.03 & \\
 28. & WISEA\,J1301$+$2807 &  110 $\pm$12.0$^{\rm c}$ &   82.6 $\pm$0.6 &  72.4 $\pm$7.2~ & \nodata        &  41.9 $\pm$0.1$^{\rm a}$ &  \nodata                  & $-$0.54 $\pm$0.02 & 3.4 $\times$ 10$^{34}$ \\
 29. & WISEA\,J1301$+$2813 &  390 $\pm$55.4$^{\rm b}$ &  165.7 $\pm$0.6 & 181.4 $\pm$6.7~ & \nodata        &  54.6 $\pm$0.3$^{\rm a}$ &  \nodata                  & $-$0.89 $\pm$0.06 & \\
 30. & WISEA\,J1301$+$2829 &  \nodata                 &   76.9 $\pm$0.7 & 242.9 $\pm$8.4$^{\rm g}$ &\nodata&  17.1 $\pm$1.0$^{\rm e}$ &  \nodata                  & $-$1.20 $\pm$0.03 & \\
 31. & WISEA\,J1301$+$2832 &  \nodata                 &  150.6 $\pm$0.9 & 242.9 $\pm$8.4$^{\rm g}$ &\nodata&  57.2 $\pm$0.4$^{\rm e}$ &  \nodata                  & $-$0.77 $\pm$0.03 & \\
 32. & WISEA\,J1302$+$2802 &  120 $\pm$12.0$^{\rm c}$ &   66.4 $\pm$0.1 &  88.5 $\pm$5.2~ & \nodata        &  25.8 $\pm$0.5$^{\rm a}$ &  \nodata                  & $-$0.75 $\pm$0.03 & \\
\hline
\end{tabular}
\label{tab:fd-spec}
\tablecomments{ The integrated flux densities quoted are in mJy along with corresponding error bars when available.  The flux densities of both extended and pointlike dominant ($S_\nu \gtrsim 30$~mJy) radio sources at 400 MHz (Col. 3) and at 1250~MHz (Col. 5) are reported.
The flux density is determined either using \textsc{aips} task \textsc{tvstat} for irregular shaped radio source or using \textsc{aips} task \textsc{jmfit} for our 250--500 MHz band and 1050--1450 band data.  The error-bar on the measurement of flux density is based on the local \textsc{rms} noise as evaluated in a circle of 5$^{\prime}$ in diameter centered on the source position.  Consequently, in the vicinity of strong sources, including the sources listed here, the local \textsc{rms} noise was sometimes higher than the \textsc{rms} noise in empty regions.
The measurements at 408~MHz (Col. 4) are from \citet{Kimetal1994}.
The rest of the measurements at 151 MHz (Col. 2), 1400 MHz (Col. 6) and 4850 MHz (Col. 7) are from NED. The spectral index (Col. 8) correspond to line representing the best fitting regression to this data. The radio luminosity (Col. 9) is evaluated at 400\,MHz for 16 sources that have known spectroscopic redshift (see Table~\ref{tab:peel-model}). \\
References: The references for flux density measurements from NED and related notes are coded as follows: (a) \citet{MHM2009} and their radio images have sensitivities ranging from 39 to 45 $\mu$Jy per 4\farcs4 beam; (b) \citet{1996MNRAS.282..779W} and they quote an error of about 10\% or less in flux densities; (c) \citet{Cordey1985} and the final map has an \textsc{rms} level of 12~mJy~beam$^{-1}$; (d) \citet{1991ApJS...75....1B} and they report an uncertainty of $\sim$4--6~mJy to the derived flux density of a source; (e) \citet{Condonetal1998} and they state that \textsc{rms} fluctuation level is 0.45~mJy~beam$^{-1}$ for total intensity images; (f) see Sec.~\ref{notes-dominant} for a discussion; (g) the measurement 242.9~mJy includes flux densities for two sources, Source\_IDs 30 and 31 \citep[see also Sec.~\ref{notes-dominant} for a discussion;][]{Kimetal1994}, and they report noise on their maps is 8~mJy~beam$^{-1}$; (h) \citet{Bonafedeetal2010} and they quote calibration error on the measured flux densities of $\sim$5\%.}
\end{table*}

\item[21.] WISEA\,J130043.86$+$282458.8 $-$
A WAT radio source with a radio core at the center, which coincides with the \textit{WISE} source.
The northern radio jet seems to be moving straight, whereas the southern jet
has been bent.  The radio morphology also suggests it to be of a ``hybrid",
with the northern jet to be of FR\,I type and southern jet to be of FR\,II
type.  Two jets show varying projected radio sizes, but
the southern jet may have encountered more cluster gas, and hence the ram
pressure, as compared to the northern jet.
The flux density at 1400 MHz = 9.4 $\pm$0.2 mJy \citep[corresponds to C1A$-$923;][]{MHM2009}, which makes spectral index $\alpha$(400--1400 MHz) = $-$0.96 $\pm$0.02.
The apparent steep spectrum could possibly be due to extended emission from this WAT source that is resolved at high frequency.
\citep{MHM2009} report that the source deblends with C1A$-$922 and possibly adjoins C1A$-$152, C1A$-$153, and C1A$-$156 as well.  \citet{Condonetal1998} note $S_{\rm 1400\,MHz}$ = 9.2 $\pm$1.1 mJy (the NVSS survey image),
which is consistent with flux density of C1A$-$923 \citep{MHM2009}.
\item[23.] SDSS\,J130055.21$+$283205.7 $-$ 
An unresolved radio source, which is $\sim$9$^{\prime\prime}$ away from the WISEA\,J130055.69$+$283200.8 (NED classification) source.  This NED classified {\it WISE} source has a photometric redshift of 0.25737.
\item[24.] NGC\,4911 $-$
The host galaxy is a very bright giant early spiral of an Sb morphological type.
The source is possibly interacting with its neighbor DRCG\,27--62 \citep{2000AJ....119..580B}.
\item[25.] WISEA\,J130106.38$+$281813.0 $-$
A classical two-sided radio galaxy showing double lobe FR\,II morphology. The radio lobe toward the west seems to be approaching us since the unresolved radio core and the associated jet are linked to it.
\item[26.] $[$HB89$]$ 1258$+$286 NED02 $-$
A marginally resolved flat spectrum ($\alpha$ = $-$0.44 $\pm$0.03) quasar source.  In addition to a bright nucleus, the source shows core-jet morphology or a weak extension in the east-west direction. It is a background source and not part of the Coma cluster of galaxies.
\item[30.] WISEA\,J130158.70$+$282928.7 $-$
A marginally resolved compact radio source.
It seems that the integrated flux density, $S_{\rm 408\,MHz}$ \citep[= 242.9 mJy;][]{Kimetal1994} is a factor of $\sim$3.2 higher than our measurement.
The \citet{Kimetal1994} measurement includes flux densities for 5C\,4.137 (Source\_ID 30) and for 5C\,4.138 (Source\_ID 31) sources, and hence their combined measurement is consistent within errors with our measurement for these two sources.
\item[31.] WISEA\,J130159.77$+$283210.1 $-$
An unresolved radio source with a possible extension toward the southeast.
As stated above (see notes on Source\_ID 30), the \citet{Kimetal1994} measurement is consistent within errors with our measurement.
\end{itemize}

\subsection{Integrated radio spectra and luminosities}
\label{rad-spec}

\begin{figure}[ht]
\begin{center}
\begin{tabular}{c}
\includegraphics[height=6.6cm]{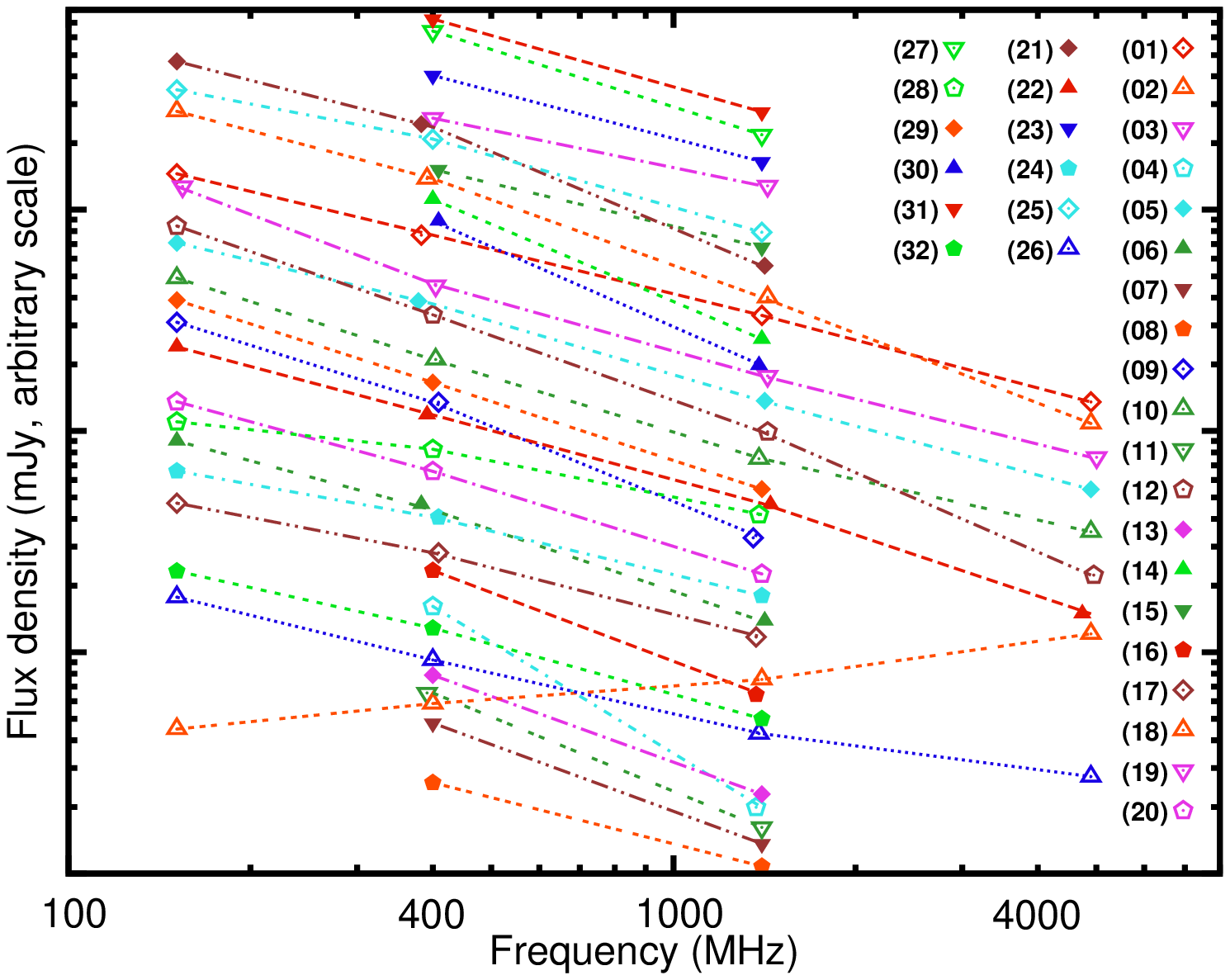}
\end{tabular}
\caption{Integrated flux densities of 32 brightest and dominant, both pointlike and extended, radio sources in the Coma cluster at the 250--500 MHz band of the uGMRT;
data at other frequencies are from the literature (see Table~\ref{tab:fd-spec} for references).  The 408~MHz data \citep{Kimetal1994} and our 1050--1450~MHz data are not plotted to avoid overcrowding.  The uGMRT flux density measurements are in good agreement with the data.  The spectra (and some of the data points) are shifted with respect to one another for clarity.}
\label{fd-spectra}
\end{center}
\end{figure}

We made comparisons of our measurements to earlier measurements, which provides a good check of our calibration.
Table~\ref{tab:fd-spec} (Cols. 2--7) lists integrated flux densities at frequencies along with error bars in the measurements and hence the integrated spectra (Col. 8, see also Fig.~\ref{fd-spectra}) for 32 brightest and dominant, both pointlike and extended, radio sources.
Qualitatively, the radio spectra are classified as (i) straight; (ii) curved, both concave and convex; and (iii) complex \citep{1980MNRAS.190..903L}.
The two mechanisms that are most likely to produce curvature in the spectrum of radiation from an electron energy distribution initially of the power-law form are \citep[e.g.][]{Miley1980}: (i) synchrotron energy losses, which cause downward curvature at high frequencies, and (ii) self-absorption of radiation from regions of high brightness temperatures, e.g, compact sources.
We see in Fig.~\ref{fd-spectra} that within the errors, all our uGMRT measurements and data from the literature fall nearly along with a `straight' power law with some hint of energy losses from synchrotron cooling.  This suggests that our flux densities measurements are consistent with the literature data, providing confidence in our data calibration and its reduction methodologies.

Almost all sources, except FBQS J1300$+$2830 (Source\_ID 18), show steep spectra, an effect attributed to the presence of extended diffuse structures.  The spectra for total flux density via. best-fitting regression to data for all sources is reported in Table~\ref{tab:fd-spec} (Col. 8).
The mean and median of these sources are $-$0.83 and $-$0.78, respectively, suggesting that $\sim$59\% sources have steep spectra ($\alpha$ $<$ $-$0.7).
Of the sources, seven sources have relatively steep spectra ($-$1.0 $>$ $\alpha$ $>$ $-$1.3) and
one source has a very steep spectrum ($\alpha$ $<$ $-$1.3).
This very steep spectrum source, WISEA J125752.93$+$280609.4 (Source\_ID 04), $\alpha$ = $-$1.59 $\pm$0.02 could be a possible candidate for a high-redshift radio galaxy \citep{Saxenaetal}.
The infrared colors are not known to deviate from the predicted relations with redshift for a standard giant elliptical galaxy spectrum.
The source (WISEA J125752.93$+$280609.4) is a 13.81 mag source in $Ks$ (NED: 2MASS extended objects, final release), makes it unlikely to be a high-redshift radio galaxy from the $K$-$z$ relation for radio galaxies \citep{LillyLongair}.

Of the sources, 17 sources have known spectroscopic redshifts (Table~\ref{tab:peel-model}).
Luminosities were evaluated at 400\,MHz for the radio emission (Col. 8, Table~\ref{tab:fd-spec}).  This frequency was chosen because very accurate flux densities are available, and contributions from compact cores are small.
The radio powers have been $K$-corrected to the rest frame of each source \citep{LalHo}.

\section{Conclusions}
\label{sec.sum-conc}

In this first paper of the series, we have presented in the preceding sections details of the observations, data reduction, and performance assessment for the 250--500 MHz band and the 1050--1450 MHz band using uGMRT for the Coma cluster of galaxies.  An image of a single field of the 1.75~deg$^2$ and 0.21~deg$^2$ areas have been presented up to the \textsc{rms} noises of $\approx$21 $\mu$Jy~beam$^{-1}$ and $\approx$13 $\mu$Jy~beam$^{-1}$ with 6\farcs3 and 2\farcs2 angular resolutions at the 250--500 MHz band and the 1050--1450 MHz band, respectively, representing the deepest uGMRT image at the 250--500 MHz band.
We also provide descriptions of radio morphologies and spectra of 32 brightest and dominant, both pointlike and extended, radio sources in the field.

We are undertaking an uGMRT study at 125--250 MHz (band-2), 250--500 MHz (band-3), and 550--850 MHz (band-4) of several clusters to make precise flux density measurements of all detected radio sources and model them, thereby aiming to provide exact estimates of the radio halo emission and statistics.
The data presented here along with a wealth of multiwavelength data available is being used to investigate the head-tail radio galaxy, NGC\,4869 (Source\_ID 02, which will form paper II of this series), to study radio luminosity function, to build a sample of ultra-steep spectrum sources, etc.

\smallskip

D.V.L. thanks the anonymous referee for his/her contributions that considerably improved the manuscript.
He also thanks Tiziana Venturi and Dave Green for useful discussions, Ishwara-Chandra C.H. for discussions on some aspects of this project, and Sushan Konar for careful reading of the manuscript.
He acknowledges the support of the Department of Atomic Energy, Government of India, under project No. 12-R\&D-TFR-5.02-0700.
We thank the staff of the GMRT who made these observations possible. The GMRT is run by the National Centre for Radio Astrophysics of the Tata Institute of Fundamental Research.
This research has made use of the NED, which is operated by the Jet Propulsion Laboratory, Caltech, under contract with the NASA, and NASA's Astrophysics Data System.

\end{document}